\documentclass[showpacs,amsmath,amssymb,10pt,aps]{revtex4}

\usepackage{graphicx,color}
\usepackage{amssymb}
\usepackage{amsfonts}
\usepackage{amsbsy}
\usepackage{bbm, bm}
\usepackage[T1]{fontenc}
\usepackage[cp1250]{inputenc}
\usepackage{epstopdf}

\newtheorem{Theorem}{Theorem}

\begin{document}

\title{Photon counting probabilities of the output field for a single-photon input}

\author{Anita Magdalena D\k{a}browska}

\address{Institute of Theoretical Physics and Astrophysics, University of Gda\'nsk, ul. Wita Stwosza 57, 80-308 Gda\'nsk, Poland}


\begin{abstract}
	We derive photon counting statistics for an output field of a single-photon wave packet interacting with a quantum system (e.g. a quantum harmonic oscillator or a two-level atom). We determine the exclusive probability densities for the output field by making use of quantum filtering theory. The quantum trajectories for continuous in time measurements of the output field (reflected and transmitted), are determined starting from a collision model and difference filtering equations. We provide analytical formulae for quantum trajectories associated with a two-dimensional stochastic counting process, we describe their structure, and give a physical interpretation to them. Moreover, we provide analytical expressions for probability densities of the times of successive photon detections for a single-photon field scattered on a two-level atom, for an arbitrary photon profile and any initial state of the atom.  
\end{abstract}



\maketitle

\section{Introduction}

	The scattering of light on a quantum system is one of core issues in quantum optics. Many efforts were put recently to describe the scattering of light prepared in $N$-photon state in a one-dimensional (1D) waveguide. The reason for this interest is the rapid development of techniques of generating and manipulating single-photon and multi-photon states of light \cite{Banaszek05,Cooper13,Yukawa13,Rybarczyk13,Scarani2013,Rempe15,Lodahl15,Ogawa16,Lodahl17,Sun18,Leong2016}. There are several theoretical techniques allowing to tackle this problem. The scattering in waveguide can be studied by using the Heisenberg picture approach \cite{Domokos02,WMSS11,WMS12,Stolyarov2013,Scarani2016a}, methods based on the Lippmann-Schwinger equation \cite{ShenFan2005,Fan2005b,ShenFan2007a,ShenFan2007b,Fan2009,Fan2012,Gritsev2012,Shen2015}, input-output formalism \cite{Fan2010,Fan2013}, generalized master equation \cite{GEPZ98,Baragiola2012,Cirac2015}, pure-state analysis \cite{Konyk16,Nysteen2015} or a stochastic approach \cite{Gough12a,Gough12b,Gough13,Dong15,Song16,Pan16,Baragiola17,Dabrowska17,Dabrowska18,Dabrowska19,Zhang19,Dong19,Dabrowska2020a}. In \cite{Scarani2016b} a method based on the operational translation of the system nonlinear response is described. In \cite{Fischer2018a, Fischer2018b} the scattering into 1D is tackled by a discrete approximation of bath Hilbert space. A comprehensive review on this subject was given in  \cite{Zubairy2016,Roy2017}.

	In this paper, we analyze the problem of scattering of a single-photon wave packet on a quantum system, within the Wesskopf-Wigner approximation \cite{Scully1997}, by making use of quantum filtering theory \cite{BarBel91,Car93,BGM04,B05,GZ10,WM10}.  We describe the properties of the output field (the field after an interaction with quantum system) in time domain. The input-output formalism and the quantum filtering theory were formulated in the framework of the quantum stochastic It\^{o} calculus  (QSC) \cite{HP84,Par92}. To find the exclusive counting probability densities \cite{BarBel91,Car93,Srinivas81,Srinivas1982,Mollow68}, which fully characterize the statistics of photons in the output field, we determine {\it quantum trajectories} \cite{Car93} --- the conditional states of the system depending on the results of continuous in time measurement performed on the output field. These states are also called {\it a posteriori} states \cite{BarBel91}. Due to the temporal correlations of the input field, the evolution of open quantum system is non-Markovian in this case  \cite{Ciccarello2018,Dabrowska2020}. It is therefore not governed by Gorini-Kossakowski-Linblad-Sudarshan master equation \cite{GKS76,Lin76} but by a set of coupled master equations. Instead of studying an averaged evolution of the quantum system, given by the set of master equations, we analyze the set of filtering equations --- the set of stochastic equations for the conditional operators of the system.


  	 We formulate the problem of scattering by making use of a discrete model of repeated interactions and measurements (collision model) \cite{ B02,BH08,BHJ09,Kretschmer2016,Breuer2016,Ciccarello2017,Gross18,M95, G04, GS04, Attal06, P08, PP09, P10,Ciccarello2021} with a bidirectional field defined as two collections of ``ancillas'' which are two-level systems (qubits). One of these collections is taken in an entangled state --- a discrete analogue of continuous-mode single photon state \cite{L00,RMS07,Milburn08}, and the second collection is prepared in the vacuum, which means that all its qubits are in the ground state.  We assume that bath qubits do not interact with each other and there is no initial correlation between the quantum system and its environment. Successive interactions (``collisions'') of the environment with the quantum system are described by unitary operators involving the system and two bath qubits. A schematic sketch of the discrete dynamics of the composed system, consisting of the travelling environment and the quantum system, is shown in Fig. \ref{Pic}. To characterize the properties of the output field, we assume that the measurements are performed on all bath elements after their interactions with the system. The results of these measurements give rise to the conditional dynamics of the quantum system, which is described by a set of difference filtering equations in the discrete case, and by a set of differential filtering equations in the continuous-time limit. Instead of solving the set differential filtering equations numerically, as it is usually done, we derive analytical formulae for the corresponding quantum trajectories. This allows us to determine general expressions for exclusive counting probability densities. We describe in detail the results for the case when a single-photon field interacts with a two-level atom. We fully characterize the statistics of photons of the output field in this case, and we show how to use the photon counting probability densities to determine the probability densities of successive counts and the mean number of photons registered up to a given time. In this paper, we present a direct generalization of the results for a single-photon unidirectional field described in \cite{Dabrowska17,Dabrowska19,Dabrowska2020a}.

	We would like to emphasize that the collision model provides the results which are consistent with the studies based on quantum stochastic calculus \cite{Gough12a,Gough12b,Baragiola2012,Baragiola17}, but the discrete approach allows one to simplify calculations and to provide an intuitive and rigorous interpretation to quantum trajectories. It allows the reader unfamiliar with QSC to understand the stochastic approach. The decomposition of the reduced system dynamics 
	into quantum trajectories helps also one to better understand the nontrivial evolution of a quantum system interacting with a temporally correlated field.

	The derivation of the Markovian case of conditional and unconditional evolution of an open quantum system in the discrete model of repeated interactions and measurements, with the bath defined as a sequence of qubits, one can find, for instance, in \cite{B02,GS04, Gross18}. For a discussions on physical assumptions leading to collision models in quantum optics see, for instance, \cite{Gross18,Ciccarello2017,Fischer2018a}.

	The paper is organized as follows. We define the collision model for a bidirectional field in Sec. II. In Sec. III, we provide a description of repeated measurements and we derive the formula for the conditional state of a system for discrete stochastic evolution. In Sec. IV, we determine the set of difference filtering equations and we obtain its continuous-time limit. Next, in Sec. V, we give analytical formulae for conditional vectors defining the quantum trajectories and we provide the expressions for photon counting probability densities in the continuous-time limit. In Sec. VI, we derive the statistics of counts for the output field for a two-level atom.

\section{Collision model. Repeated interactions for a bidirectional field}

\begin{figure}
	\includegraphics[width=7cm]{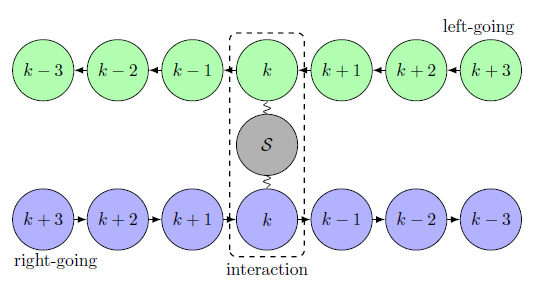}
	\caption{The system $\mathcal{S}$ interacts with a bidirectional field: the right-going single-photon pulse $|1_\xi\rangle$ and the left-going vacuum.} \label{Pic}
\end{figure}

We consider a quantum system $\mathcal{S}$ interacting with an environment $\mathcal{E}$ being a bidirectional field modeled by two chains of qubits. One chain describes the field going to the right and the second one refers to the field going to the left. We assume that the qubits do not interact with each other but qubits of each chain interact subsequently with the system $\mathcal{S}$. At a given moment $\mathcal{S}$ interacts with only two qubits: 
one coming from the left and the other one coming from the right. Any interaction (``collision'') has the same duration $\tau$ and each of the bath qubits interacts with the system only once. The schematic of the collision dynamics is shown in Fig. \ref{Pic}. We describe the dynamics of the composed system $\mathcal{E}+\mathcal{S}$  up to time $T=N\tau$, where $N$ denotes the number of qubits in each chain. The Hilbert space of the environment is then defined as
\begin{equation}
\mathcal{H}_{\mathcal{E}}=\mathcal{H}_{\mathcal{E}_{1}}\otimes\mathcal{H}_{\mathcal{E}_{2}},
\end{equation}
where
\begin{equation}
\mathcal{H}_{\mathcal{E}_{l}}=\bigotimes_{k=0}^{N-1}\mathcal{H}_{\mathcal{E}_{l,k}},\;l=1,2
\end{equation}
and $\mathcal{H}_{\mathcal{E}_{l,k}}=\mathbb{C}^2$ is the Hilbert space of the qubit of the $l$-th part of the environment which interacts with $\mathcal{S}$ in the time interval $[k\tau, (k+1)\tau)$. Note that the Hilbert spaces $\mathcal{H}_{\mathcal{E}_{l}}, \;l=1,2$ can be split as
\begin{equation}
\mathcal{H}_{\mathcal{E}_{l}}=\mathcal{H}_{\mathcal{E}_{l}}^{j-1]}\otimes \mathcal{H}_{\mathcal{E}_{l}}^{[j},
\end{equation}
where
\begin{equation}
\mathcal{H}_{\mathcal{E}_{l}}^{j-1]}=\bigotimes_{k=0}^{j-1}\mathcal{H}_{\mathcal{E}_{l,k}},\;\;\;
\mathcal{H}_{\mathcal{E}_{l}}^{[j}=
\bigotimes_{k=j}^{N-1}\mathcal{H}_{\mathcal{E}_{l,k}}.
\end{equation}
Thus, if $j\tau$ is a current moment, then 
\begin{equation}
\mathcal{H}_{\mathcal{E}}^{j-1]}=\mathcal{H}_{\mathcal{E}_{1}}^{j-1]}\otimes \mathcal{H}_{\mathcal{E}_{2}}^{j-1]}
\end{equation}
refers to the part of the environment which has already interacted with $\mathcal{S}$, constituting the output field, and 
\begin{equation}
\mathcal{H}_{\mathcal{E}}^{[j}=\mathcal{H}_{\mathcal{E}_{1}}^{[j}\otimes \mathcal{H}_{\mathcal{E}_{2}}^{[j}
\end{equation}
refers to the part which has not interacted with $\mathcal{S}$ yet\textemdash the input field. We shall call $\mathcal{H}_{\mathcal{E}}^{j-1]}$ and $\mathcal{H}_{\mathcal{E}}^{[j}$ respectively {\it the past } and {\it future environment spaces}. 

The evolution of the composed system is described by the sequence of unitary operators, $U_{j}$ for $0\leq j\leq N-1$, defined by
\begin{equation}
U_{j\tau} = {V}_{j-1} {V}_{j-2} \ldots {V}_{0},\;\;\;\;\;U_{0}=\mathbbm{1},
\end{equation}
where $V_{k}$ for $0\leq k\leq N-1$ describes the interaction between $\mathcal{S}$ and $\mathcal{E}$ in the time-interval
$[k\tau, (k+1)\tau)$. The operator ${V}_{k}$ acts non-trivially only in the space $\mathcal{H}_{\mathcal{E}_{1},k}\otimes\mathcal{H}_{\mathcal{E}_{2},k}\otimes \mathcal{H}_{\mathcal{S}}$, where $\mathcal{H}_{\mathcal{S}}$ is the Hilbert space of $\mathcal{S}$, and $V_{k}$ has the form
\begin{equation}\label{Vk}
{V}_{k}=\exp\left(-i \tau H_{k}\right)
\end{equation}
with the Hamiltonian \cite{G04,Fischer2018a,Gross18,Ciccarello2017}
\begin{eqnarray}\label{hamiltonian} \
H_{k}=H_{\mathcal{S}}+\sum_{l=1}^{2}\frac{i}{\sqrt{\tau}}\left(\sigma_{l,k}^{+}\otimes L_{l}-\sigma_{l,k}^{-}\otimes L^{\dagger}_{l}\right).
\end{eqnarray}
  We set $\hbar=1$ throughout the paper and to simplify the notation we omit a multiplication by identity operators. The model is formulated in the framework of some standard assumptions made in quantum optics: rotating wave-approximation, a flat coupling constant, and the extension of the lower limit of integration over frequency to minus infinity    
 \cite{Scully1997}. The bandwidth of the spectrum is assumed to be much smaller that the central frequency of the pulse. The Hamiltonian $H_{k}$ is written in the interaction picture eliminating the free evolution of the field.  Here $H_\mathcal{S}$ stands for the Hamiltonian of $\mathcal{S}$,  $\sigma^{+}_{l,k}$ and $\sigma^{-}_{l,k}$ denote respectively the raising and lowering operators acting in $\mathcal{H}_{\mathcal{E}_{l,k}}$. From the mathematical point of view, $L_{l}$ for $l=1,2$ are arbitrary bounded operators on $\mathcal{H}_S$. They are called jump operators or the It\^{o} coefficients. In Sec. V we consider $\mathcal{S}$ which is a two-level atom and we define $L_{l}$ as $\sqrt{\Gamma_{l}}\sigma_{-}$, where $\Gamma_{l}$ is a positive coupling constant and $\sigma_{-}$ is the atom lowering operator. If $\mathcal{S}$ is a two-sided cavity, then  $L_{l}=\sqrt{\Gamma_{l}}a$ and $a$ is the annihilation operator of a cavity mode. A discussion about the structure of the collision model and physical assumptions leading to it one can find, for instance, in
 	\cite{Gross18,Fischer2018a,Ciccarello2017}. We use the representation of (\ref{Vk}) in the basis $\{|00\rangle_{k}, |01\rangle_{k}, |10\rangle_{k}, |11\rangle_{k}\}$, such that
\begin{equation}\label{operatorV}
\exp\left(-i\tau H_{k}\right)  = \sum_{i_{1},i_{2},i_{3},i_{4}=0,1} |i_{1}i_{2}\rangle_{k} {}_{k}\langle i_{3}i_{4}| \otimes V_{i_{1}i_{2},i_{3}i_{4}}
\end{equation}
where $|i_{l_{1}}i_{l_{2}}\rangle_{k}=|i_{l_{1}}\rangle_{k}\otimes |i_{l_{2}}\rangle_{k}$, $V_{i_{1}i_{2},i_{3}i_{4}}$ are operators on $\mathcal{S}$, and their explicit forms are given in \ref{V}. 
Clearly, in order to approximate the continuous evolution, we take $\tau$ that satisfies the conditions
	\begin{equation}
	\tau\ll \Gamma_{l}^{-1},\;\; l=1,2.
	\end{equation}
	Note that we are interested in terms up to order $\tau$ for determining first-order differential equations, it means that we have to expand $V_{k}$ to the second order. To obtain the continuous dynamics, we will take finally the limit of $\tau\to 0$ and $N\to \infty$ such that $T=N\tau$ is fixed.

The initial state of the composed system is assumed to be the product state vector of the form
\begin{equation}\label{ini0}
|\Psi_{0}\rangle=|1_{\xi}\rangle\otimes |vac\rangle\otimes|\psi_{0}\rangle,
\end{equation}
where $|\psi_{0}\rangle$ is the initial state of $\mathcal{S}$ and
\begin{eqnarray}
|1_{\xi}\rangle=\sum_{k=0}^{N-1}\sqrt{\tau}\xi_{k}\sigma^{+}_{1,k}|vac\rangle
\end{eqnarray}
with the vacuum vector $|vac\rangle=|0\rangle_{0}\otimes |0\rangle_{1}\otimes \ldots\otimes|0\rangle_{N-1}$, where $|0\rangle_{k}$ is the ground state in $\mathbb{C}^2$, and $\displaystyle{\sum_{k=0}^{N-1}}\tau|\xi_{k}|^2 = 1$.
Note that vector 
$|1_{\xi}\rangle$ poses the additive decomposition property
\begin{equation}
|1_{\xi}\rangle
=  \sum_{k=0}^{j}\sqrt{\tau} \xi_{k}\sigma^{+}_{1,k}|vac\rangle +   \sum_{k=j+1}^{N-1}\sqrt{\tau} \xi_{k}\sigma^{+}_{1,k}|vac\rangle
\end{equation}
and it can be written in the form
\begin{equation}|1_{\xi}\rangle =    \sum_{k=0}^{N-1}\sqrt{\tau} \xi_{k}|1_k \rangle,
\end{equation}
where $|1_k \rangle=|0\rangle_{0}\otimes |0\rangle_{1} \otimes\ldots |0\rangle_{k-1}\otimes|1\rangle_{k}\otimes |0\rangle_{k+1}\otimes \ldots|0\rangle_{N-1}$. Thus $|1_{\xi}\rangle$ is a superposition of vectors with one qubit prepared in the excited state and all the others in the ground state. Clearly, $|1_{\xi}\rangle$ an entangled state and  $|\xi_{k}|^2\tau$ is the probability that the qubit of the $k$ number in the first chain is in the excited state and all the others qubits in this chain are in the ground state. One can easily check the identities
\begin{equation}
\sigma^{-}_{1,k}|1_{\xi}\rangle = \sqrt{\tau}\xi_{k}|vac\rangle,
\end{equation}
\begin{equation}
\sigma^{+}_{1,k}\sigma^{-}_{1,k}|1_{\xi}\rangle = \sqrt{\tau}\xi_{k}|1_{k}\rangle.
\end{equation}
The state $|1_{\xi}\rangle$ is a discrete version of a continuous-mode single-photon state \cite{L00, RMS07, Milburn08}. 

After $j$ interactions the state of the composed system is given as 
\begin{equation}
U_{j-1}|\Psi_{0}\rangle\langle\Psi_{0}|U_{j-1}^{\dagger}. 
\end{equation}
And taking the partial trace over the environment, we obtain the reduced state of $\mathcal{S}$ at time $j\tau$:
\begin{equation}\label{apriori}
\varrho_{j}=\mathrm{Tr}_{\mathcal{E}}\left[U_{j-1}|\Psi_{0}\rangle\langle\Psi_{0}|U_{j-1}^{\dagger}\right]. 
\end{equation}

\section{Repeated measurements and conditional state}

We describe now the setup of repeated measurements performed on the environment chains. We assume that after each step of interaction, the measurements are performed on two qubits which have just interacted with $\mathcal{S}$. Clearly, the first chain is going to the right and its output is measured on the right side of $\mathcal{S}$, the second chain is going to the left and its output is measured on the left side of $\mathcal{S}$. We consider measurements of the input observables:
\begin{equation}\label{obs}
\sigma_{l,k}^{+}\sigma_{l,k}^{-},\;\;l=1,2,\; k=0, 1, \ldots N-1.
\end{equation}
It means that we study the measurements in the orthogonal basis: $\{|00\rangle_{k}, |01\rangle_{k}, |10\rangle_{k},$ $|11\rangle_{k}\}$. 

One can check that after the first interaction, the state of the composed system has the following form
\begin{eqnarray}
&&\lefteqn{{V}_{0}|\Psi_{0}\rangle=|vac\rangle\otimes |vac\rangle\otimes (-\tau \xi_{0} L_{1}^{\dagger})|\psi_{0}\rangle}\nonumber\\
&&+|0\rangle_{0}\otimes \sum_{k=1}^{N-1}\sqrt{\tau}\xi_{k}|1_{k}\rangle_{[1}\otimes |vac\rangle\otimes \left(1-i\tau H_{\mathcal{S}}-\frac{\tau}{2}\left(L_{1}^{\dagger}L_{1}+L_{2}^{\dagger}L_{2}\right)\right)|\psi_{0}\rangle\nonumber\\
&&+|1\rangle_{0}\otimes\sum_{k=1}^{N-1}\sqrt{\tau}\xi_{k}|1_{k}\rangle_{[1}\otimes |vac\rangle\otimes \sqrt{\tau}L_{1}|\psi_{0}\rangle\nonumber\\
&&+|1\rangle_{0}\otimes|vac\rangle_{[1}\otimes |vac\rangle\otimes \sqrt{\tau}\xi_{0}|\psi_{0}\rangle
\nonumber\\
&&+|0\rangle_{0}\otimes \sum_{k=1}^{N-1}\sqrt{\tau}\xi_{k}|1_{k}\rangle_{[1}\otimes |1\rangle_{0}\otimes|vac\rangle_{[1,}\otimes \sqrt{\tau}L_{2}|\psi_{0}\rangle
+\nonumber\\
&&+|1\rangle_{0}\otimes |vac\rangle_{[1}\otimes|1\rangle_{0}\otimes |vac\rangle_{[1}\otimes \tau \xi_{0}L_{2}|\psi_{0}\rangle 
+\nonumber\\
&&+|1\rangle_{0}\otimes \sum_{k=1}^{N-1}\sqrt{\tau}\xi_{k}|1_{k}\rangle_{[1}\otimes|1\rangle_{0}\otimes |vac\rangle_{[1}\otimes \frac{\tau}{2} \left(L_{1}L_{2}+L_{2}L_{1}\right)|\psi_{0}\rangle+O(\tau^{3/2}),
\end{eqnarray}    
where
\begin{equation}
|vac\rangle_{[1}=|0\rangle_{1}\otimes |0\rangle_{2} \otimes  \ldots\otimes|0\rangle_{N-1},
\end{equation}
\begin{equation}
|1_k \rangle_{[1}=|0\rangle_{1}\otimes |0\rangle_{2} \otimes\ldots |0\rangle_{k-1}\otimes|1\rangle_{k}\otimes |0\rangle_{k+1}\otimes \ldots|0\rangle_{N-1},
\end{equation}
and $O(.)$ is the Landau symbol. We assume that now the measurement is performed in the  basis $\{|00\rangle_{0}, |01\rangle_{0}, |10\rangle_{0}, |11\rangle_{0}\}$. 
Thus, if we have observed, for instance, the result $(0,0)$ (zero counts for the both detectors), then the conditional state of the composed system is given by
\begin{equation}
\frac{|00\rangle_{0}\langle 00|{V}_{0}|\Psi_{0}\rangle\langle\Psi_{0}|{V}_{0}^{\dagger}|00\rangle_{0}\langle 00|}{\mathrm{Tr}_{\mathcal{E+S}}\left[{V}_{0}|\Psi_{0}\rangle\langle\Psi_{0}|{V}_{0}^{\dagger}|00\rangle_{0}\langle 00|\right]}. 
\end{equation}
 Note that it is a pure state. By eliminating the degrees of freedom of the past environment, which will not interact with $\mathcal{S}$ in the future, we obtain the recipe for the {\it a posteriori} state of $\mathcal{S}$ and the input part of the environment. Thus the conditional state of $\mathcal{S}$ and the input part of the environment at time $\tau$ can be written in the form
\begin{equation}
|\tilde{\Psi}_{1| (\eta_{1,1},\eta_{2,1})}\rangle = \frac{|\Psi_{1| (\eta_{1,1},\eta_{2,1})} \rangle}{\sqrt{\langle\Psi_{1| (\eta_{1,1},\eta_{2,1})}|\Psi_{1| (\eta_{1,1},\eta_{2,1})}\rangle}},
\end{equation}
where $(\eta_{1,1},\eta_{2,1})$ stands for the results of the first measurement performed on the qubits of two chains at time $\tau$. Clearly, $|\Psi_{1| (\eta_{1,1},\eta_{2,1})} \rangle$ is a random unnormalized vector from the Hilbert space $\mathcal{H}_{\mathcal{E}}^{[1}\otimes \mathcal{H_S}$ and its form depends on the observed outcomes. For instance, if the result is $(0,0)$, then we get
\begin{eqnarray}
|\Psi_{1| (0,0)} \rangle&=& \sum_{k=1}^{N-1}\sqrt{\tau}\xi_{k}|1_{k}\rangle_{[1}\otimes |vac\rangle_{[1}\otimes \left(1-i\tau H_{\mathcal{S}}-\frac{\tau}{2}\left(L_{1}^{\dagger}L_{1}+L_{2}^{\dagger}L_{2}\right)\right)|\psi_{0}\rangle\nonumber\\
&&-|vac\rangle_{[1}\otimes |vac\rangle_{[1}\otimes \tau \xi_{0} L_{1}^{\dagger}|\psi_{0}\rangle.
\end{eqnarray}
One can easily check that the results of the first measurement appear with the following probabilities
\begin{equation}
p_{1}\left((0,0)\right)=1-\tau\left(\langle\psi_{0}|(L_{1}^{\dagger}L_{1}+L_{2}^{\dagger}L_{2})|\psi_{0}\rangle+|\xi_{0}|^2\right)+O(\tau^2),
\end{equation}  
\begin{equation}
p_{1}\left((1,0)\right)=\tau\left(\langle\psi_{0}|L_{1}^{\dagger}L_{1}|\psi_{0}\rangle+|\xi_{0}|^2\right)+O(\tau^2),
\end{equation}
\begin{equation}
p_{1}\left((0,1)\right)=\tau\langle\psi_{0}|L_{2}^{\dagger}L_{2}|\psi_{0}\rangle+O(\tau^2),
\end{equation}
\begin{equation}
p_{1}\left((1,1)\right)=O(\tau^2).
\end{equation} 
Since the probability of the result $(1,1)$ is $O(\tau^2)$, we ignore such detection. In the next step the system $\mathcal{S}$ interacts with the second pair of the qubits and after this interaction we perform the next measurement. Note that the conditional state of $\mathcal{S}$ and the input field in time $2\tau$ will depend on the results of the two past measurements.  
 We formulate our result for time $j\tau$ $(1\leq j\leq N-1)$ in the form of a theorem. 

\begin{Theorem}\label{TH-1}  The conditional state of $\mathcal{S}$ and the input part of the environment, which has not interacted with $\mathcal{S}$ up to $j\tau $, for the initial state (\ref{ini0}) and the measurements of (\ref{obs}) is at time $j\tau $ given by
	\begin{equation}\label{cond1}
	|\tilde{\Psi}_{j| \pmb{\eta}_j}\rangle = \frac{|\Psi_{j| \pmb{ \eta}_j}\rangle}{\sqrt{\langle\Psi_{j| \pmb{\eta}_j}|\Psi_{j| \pmb{\eta}_j}\rangle}},
	\end{equation}
	where $ |\Psi_{j|\pmb{\eta}_j} \rangle $ is the unnormalized conditional vector from the Hilbert space $\mathcal{H}_{\mathcal{E}}^{[j} \otimes \mathcal{H}_\mathcal{S}$ having the form
	\begin{eqnarray}\label{cond2}
	|\Psi_{j|\pmb{\eta}_j} \rangle =   \sum_{k=j}^{N-1}\sqrt{\tau} \xi_{k}|1_{k}\rangle_{[j} \otimes | vac \rangle_{[j} \otimes |\alpha_{j| \pmb{\eta}_j}\rangle +| vac \rangle_{[j} \otimes | vac \rangle_{[j} \otimes |\beta_{j| \pmb{\eta}_j}\rangle
	\end{eqnarray}
	where 
	\begin{equation}
	|vac\rangle_{[j}=|0\rangle_{j}\otimes |0\rangle_{j+1} \otimes |0\rangle_{2} \ldots|0\rangle_{N-1},
	\end{equation}
	\begin{equation}
	|1_k \rangle_{[j}=|0\rangle_{j}\otimes |0\rangle_{j+1} \otimes\ldots |0\rangle_{k-1}\otimes|1\rangle_{k}\otimes |0\rangle_{k+1}\otimes \ldots|0\rangle_{N-1},
	\end{equation}
	and $\pmb{\eta}_j$ is a $j$-vector $\pmb{\eta}_j = (\eta_{j},\eta_{j-1},\ldots,\eta_1)$ with $\eta_k= (\eta_{1,j},\eta_{2,j})$, and $\eta_{l,j}= \{0,1\}$ for $l=1,2$. The $\pmb{\eta}_j$ represents results of all measurements of (\ref{obs}) up to time $j\tau$ and
	the elements of the pairs denote respectively results of the measurements performed on qubits of the first and the second chain. 
	The vectors $|\alpha_{j| \pmb{\eta}_j}\rangle$, $|\beta_{j| \pmb{\eta}_j}\rangle$, from the Hilbert space $\mathcal{H}_{\mathcal{S}}$, satisfy the respective recurrence equations:
	\begin{enumerate}
	\item for $\eta_{j+1}=(0,0)$
	\begin{eqnarray}\label{rec1a}
	|\alpha_{j+1| \pmb{\eta}_{j+1} }\rangle &=& V_{00, 00} |\alpha_{j| \pmb{\eta}_j}\rangle,\\
	\label{rec1b}
	|\beta_{j+1| \pmb{\eta}_{j+1} }\rangle &=& V_{00,00} |\beta_{j| \pmb{\eta}_j}\rangle + \sqrt{\tau} \xi_{j} V_{00, 10} |\alpha_{j| \pmb{\eta}_j}\rangle, 
	\end{eqnarray}
		\item for $\eta_{j+1}=(1,0)$
	\begin{eqnarray}\label{rec2a}
	|\alpha_{j+1| \pmb{\eta}_{j+1} }\rangle &=& V_{10, 00} |\alpha_{j| \pmb{\eta}_j}\rangle,\\
	\label{rec2b}
	|\beta_{j+1| \pmb{\eta}_{j+1} }\rangle &=& V_{10,00} |\beta_{j| \pmb{\eta}_j}\rangle + \sqrt{\tau} \xi_{j} V_{10, 10} |\alpha_{j| \pmb{\eta}_j}\rangle, 
	\end{eqnarray}
	\item for $\eta_{j+1}=(0,1)$
	\begin{eqnarray}\label{rec3a}
	|\alpha_{j+1| \pmb{\eta}_{j+1} }\rangle &=& V_{01, 00} |\alpha_{j| \pmb{\eta}_j}\rangle,\\
	\label{rec3b}
	|\beta_{j+1| \pmb{\eta}_{j+1} }\rangle &=& V_{01,00} |\beta_{j| \pmb{\eta}_j}\rangle+\sqrt{\tau}\xi_{j}V_{01,10}|\alpha_{j| \pmb{\eta}_j}\rangle. 
	\end{eqnarray}
	\item for $\eta_{j+1}=(1,1)$
	\begin{eqnarray}\label{rec4a}
	|\alpha_{j+1| \pmb{\eta}_{j+1} }\rangle &=& V_{11, 00} |\alpha_{j| \pmb{\eta}_j}\rangle,\\
	\label{rec4b}
	|\beta_{j+1| \pmb{\eta}_{j+1} }\rangle &=& V_{11,00} |\beta_{j| \pmb{\eta}_j}\rangle+\sqrt{\tau}\xi_{j}V_{11,10}|\alpha_{j| \pmb{\eta}_j}\rangle.	
	\end{eqnarray}
\end{enumerate}
	Initially we have $|\alpha_{j=0}\rangle=|\psi_{0}\rangle$, $|\beta_{j=0}\rangle=0$. 
\end{Theorem}
The proof of Theorem \ref{TH-1} one can find in \ref{proof}.

Note that the vectors $|\alpha_{j| \pmb{\eta}_j}\rangle$, $|\beta_{j| \pmb{\eta}_j}\rangle$ depend on all results of measurements up to the time $j\tau$ and this is indicated by their subscript $\pmb{\eta}_j$.  It is seen from the formula (\ref{cond2}) that the system $\mathcal{S}$ becomes entangled with the left input part of the environment. This property distinguishes the studied case from the Markovian one within after each measurement the state of the system and the part of environment which has not interacted with the system is a product state (see, for instance, \cite{G04}). Clearly, the non-Markovian character of the evolution comes here from the initial entanglement of the bath qubits. Let us notice that the conditional state (\ref{cond1}) has the following physical interpretation. The first term (with the conditional vector $|\alpha_{j| \pmb{\eta}_j}\rangle$) describes the scenario that $\mathcal{S}$ has not met the qubit prepared in the upper state yet and it appears in the future. The second term (with the vector $|\beta_{j| \pmb{\eta}_j}\rangle$) is associated with the scenario that $\mathcal{S}$ has already interacted with the qubit prepared in the upper state and it meets in the future only qubits in the ground states. We treat $\tau$ as a small time step in the sense that all elements of the recurrence equations (\ref{rec1a})-(\ref{rec4b}) of order more than $\tau$ can be ignored. To understand why these terms disappear one needs to consider the continuous-time limits of the solutions to the derived equations.

The probability of a given sequence, $\pmb{\eta}_j$, is
\begin{equation}
p(\pmb{\eta}_j)=\langle {\Psi}_{j| \pmb{\eta}_j}|{\Psi}_{j| \pmb{\eta}_j}\rangle
\end{equation}
and one can easily check that it can be expressed by the conditional vectors as 
\begin{equation}
p(\pmb{\eta}_j)= \langle\alpha_{j|  \pmb{\eta}_{j}}|\alpha_{j|  \pmb{\eta}_{j}}\rangle\;\displaystyle{\sum_{k=j}^{N-1}}\;\tau|\xi_{k}|^2 +
\langle\beta_{j|  \pmb{\eta}_{j}}|\beta_{j|  \pmb{\eta}_{j}}\rangle.
\end{equation}

Let us note that the conditional probability of the outcome $\eta_{j+1}=(0,0)$ at $(j+1)\tau$ given the sequence $\pmb{\eta}_{j}$ is defined as
\begin{equation}\label{conpro}
p_{j+1}\left((0,0)|\pmb{\eta}_j\right)=\frac{\langle {\Psi}_{j| \left((0,0),\pmb{\eta}_j\right)}|{\Psi}_{j| \left((0,0),\pmb{\eta}_j\right)}\rangle}{\langle {\Psi}_{j| \pmb{\eta}_j}|{\Psi}_{j| \pmb{\eta}_j}\rangle}.
\end{equation}
In a similar way we define the conditional probabilities of the other outcomes. By using of the difference equations (\ref{rec1a})-(\ref{rec4b}), one can check the following important properties
\begin{equation}
p_{j+1}\left((0,0)|\pmb{\eta}_j\right)= 1+O(\tau),
\end{equation}
\begin{equation}
p_{j+1}\left((1,0)|\pmb{\eta}_j\right)=O(\tau),
\end{equation}
\begin{equation}
p_{j+1}\left((0,1)|\pmb{\eta}_j\right)=O(\tau),
\end{equation}
\begin{equation}
p_{j+1}\left((1,1)|\pmb{\eta}_j\right)=O(\tau^2).
\end{equation} 
They mean that most of the time we observe two vacuums and from time to time we observe a count on the left or a count on the right.  
However, because the simultaneous counts at the left and at the right detectors appear with the probability of order $O(\tau^2)$, we do not observe such result. 

Note that the $\eta_{j}$, describing the $j$-th results, is a two-dimensional random variable which is statistically dependent on the sequence $\eta_{1}, \ldots, \eta_{j-1}$. Let us introduce the two-dimensional discrete stochastic process
\begin{equation}\label{sp}
n_{j}=\left(n_{1,j},n_{2,j}\right)
\end{equation}
where 
\begin{equation}
n_{1,j}=\sum_{k=1}^{j}\eta_{1,j}, \;\; n_{2,j}=\sum_{k=1}^{j}\eta_{2,j} 
\end{equation}
are the stochastic processes referring to counts registered respectively by the right and the left detector. 

Taking the partial trace of $|\tilde{\Psi}_{j| \pmb{ \eta}_j}\rangle\langle\tilde{\Psi}_{j| \pmb{ \eta}_j}|$ over the environment, one obtains the {\it a posteriori} state of $\mathcal{S}$ at the time $j\tau$,
\begin{equation}\label{condS}
\tilde{\rho}_{j|  \pmb{\eta}_{j}}
=\frac{\rho_{j|  \pmb{\eta}_{j}}}{\mathrm{Tr}\rho_{j|  \pmb{\eta}_{j}}},
\end{equation}
where
\begin{equation}
\rho_{j|  \pmb{\eta}_{j}} = |\alpha_{j|  \pmb{\eta}_{j}}\rangle\langle\alpha_{j|  \pmb{\eta}_{j}}|\;\displaystyle{\sum_{k=j}^{N-1}}\;\tau|\xi_{k}|^2 +
|\beta_{j|  \pmb{\eta}_{j}}\rangle\langle\beta_{j|  \pmb{\eta}_{j}}|.
\end{equation}
Clearly, $\tilde{\rho}_{j|  \pmb{\eta}_{j}}$ is the state of $\mathcal{S}$ depending on the results of all measurements performed on the output field up to $j\tau$. Thus the proceeding of the repeated interactions and measurements allows us to derive discrete quantum trajectories for $\mathcal{S}$ associated with the two-dimensional counting stochastic process (\ref{sp}). Note that a different realization of the stochastic process (\ref{sp}) means a different quantum trajectory given by (\ref{condS}). 


Considering the counting stochastic process (\ref{sp}) it is convenient to replace the notation $\pmb{\eta}_j$ of all results from $0$ to $j\tau$ by writing only the location of counts in $(\eta_j,\ldots,\eta_1)$. So the sequence 
\begin{equation}
\left(D_{m},l_{m};D_{m-1},l_{m-1};\ldots;D_{1},l_{1}\right)
\end{equation}
means the following scenario: the first photon was counted at time $l_{1}\tau$ at the detector $D_1$, the second photon at time $l_{2}\tau$ at the detector $D_2$ and so on, where  $l_1<l_2<\ldots<l_{m}$, and no other photons were detected from $0$ to $j\tau$. We shall use the notation $R$ and $L$ to indicate respectively the right and left detector, thus $D_{1}, D_2, \ldots, D_{m}=R,L$. 

The discrete conditional vectors, being the solutions to Eqs. (\ref{rec1a})-(\ref{rec3b}) for some chosen sequences of the outcomes are given in \ref{condivectors}.

	By taking the average over all possible outcomes one gets from (\ref{condS}) the formula for the {\it a priori} state of the system $\mathcal{S}$:
\begin{equation}\label{mastersol}
\varrho_{j}=\rho_{j|\pmb{0}_{j}}+\sum_{m=1}^{j}\sum_{l_{m}=m}^{j}\sum_{l_{m-1}=m-1}^{j-1}\ldots\sum_{l_{1}=1}^{l_{2}-1}\sum_{D_{m},\ldots, D_{2},D_{1}=R,L}\rho_{j|D_{m},l_{m};D_{m-1},l_{m-1};\ldots;D_{1},l_{1}},
\end{equation}
where 
\begin{equation}
\rho_{j|\pmb{0}_{j}}=|\alpha_{j|\pmb{0}_{j}}\rangle\langle\alpha_{j|\pmb{0}_{j}}| \sum_{k=j}^{N-1}\tau|\xi_{k}|^2+|\beta_{j|\pmb{0}_{j}}\rangle\langle\beta_{j|\pmb{0}_{j}}|
\end{equation}
and
\begin{eqnarray}
\rho_{j|D_{m},l_{m};D_{m-1},l_{m-1};\ldots;D_{1},l_{1}}&=&|\alpha_{j|D_{m},l_{m};D_{m-1},l_{m-1};\ldots;D_{1},l_{1}}\rangle\langle\alpha_{j|D_{m},l_{m};D_{m-1},l_{m-1};\ldots;D_{1},l_{1}}| \sum_{k=j}^{N-1}\tau|\xi_{k}|^2\nonumber \\&&+|\beta_{j|D_{m},l_{m};D_{m-1},l_{m-1};\ldots;D_{1},l_{1}}\rangle\langle\beta_{j|D_{m},l_{m};D_{m-1},l_{m-1};\ldots;D_{1},l_{1}}|.
\end{eqnarray}

Clearly, for the moment $j\tau$ there are $4^{j}$ of different scenarios (trajectories), but some of them occur with zero probability. The {\it a priori} state specifies the unconditional evolution of $\mathcal{S}$. We constructed a decomposition (unraveling) of the reduced density operator of $\mathcal{S}$ defined by (\ref{apriori}).
Let us stress that for different choices of the measured observables one gets different quantum trajectories (conditional evolution of $\mathcal{S}$) but all of them lead to the same average dynamics. 

\section{Filtering equations for a bidirectional field}\label{filtering}

In this section, we shall determine the set stochastic equations defining the {\it a posteriori} evolution of $\mathcal{S}$. To simplify the notation, we skip in this section the conditional subscript.

First, we derive the recurrence formulae for a {\it a posteriori} state (\ref{condS}) to the order of $\tau$. To determine the conditional formulae, we make use of the recurrence equations (\ref{rec1a})-(\ref{rec3b}). If the results of measurements at time $(j+1)\tau$ is $\eta_{j+1}=(0,0)$, then we get
\begin{eqnarray}\label{filter1}
\rho_{j+1} &=& \rho_{j}-i[H_{\mathcal{S}},\rho_{j}]\tau-\frac{1}{2}\{L^{\dagger}_1L_1+L^{\dagger}_2L_2,\rho_{j}\}\tau\nonumber\\&&-|\beta_{j}\rangle\langle\alpha_{j}|L_1\xi_{j}^{\ast}\tau-L_1^{\dagger}|\alpha_{j}\rangle\langle\beta_{j}|\xi_{j}\tau
\nonumber\\&&- |\alpha_{j}\rangle\langle\alpha_{j}||\xi_{j}|^2\tau + O(\tau^2) ,
\end{eqnarray}
\begin{eqnarray}
|\alpha_{j+1}\rangle\langle\beta_{j+1}|
&=& |\alpha_{j}\rangle\langle\beta_{j}| - i[H_{\mathcal{S}},|\alpha_{j}\rangle\langle\beta_{j}|]\tau\nonumber\\&&
-\frac{1}{2}\{L^{\dagger}_1L_1+L^{\dagger}_2L_2,|\alpha_{j}\rangle\langle\beta_{j}|\}\tau\nonumber\\&&-|\alpha_{j}\rangle\langle\alpha_{j}|L_1\xi_{j}^{\ast}\tau + O(\tau^2) ,
\end{eqnarray}
\begin{eqnarray}
|\alpha_{j+1}\rangle\langle\alpha_{j+1}| &=& |\alpha_{j}\rangle\langle\alpha_{j}|-i[H_{\mathcal{S}},|\alpha_{j}\rangle\langle\alpha_{j}|]\tau\nonumber\\&&-\frac{1}{2}\{L^{\dagger}_1L_1+L^{\dagger}_2L_2,|\alpha_{j}\rangle\langle\alpha_{j}|\}\tau + O(\tau^2), 
\end{eqnarray}
where $\{a,b\}=ab+ba$. Let us notice that the unnormalized {\it a posteriori} operator $\rho_{j+1}$ depends on the results of the $(j+1)$-st measurements and the conditional operators: $\rho_{j}$, $|\alpha_{j}\rangle\langle\beta_j|$, $|\beta_{j}\rangle\langle\alpha_j|$, and $|\alpha_{j}\rangle\langle\alpha_j|$. The conditional probability of the outcome $(0,0)$ at time $(j+1)\tau$ given the {\it a posteriori} state $\tilde{\rho}_{j}$ is defined as
\begin{equation}
p_{j+1}\left((0,0)|\tilde{\rho}_{j}\right)=\frac{\mathrm{Tr}{\rho}_{j+1}}{\mathrm{Tr}{\rho}_{j}}.
\end{equation}
One can check that 
\begin{equation}
p_{j+1}\left((0,0)|\tilde{\rho}_{j}\right)=1-k_j\tau+O(\tau^2),
\end{equation}
where
\begin{eqnarray}\label{intensity1}
k_{j} &=& \mathrm{Tr}\left[ \left(L^{\dagger}_1L_1+L^{\dagger}_2L_2\right)\tilde{\rho}_{j}
+ \xi_{j}^{\ast}L_1|\tilde{\beta}_{j}\rangle\langle\tilde{\alpha}_{j} |
\right.\nonumber\\&&\left.+  \xi_{j} \,
|\tilde{\alpha}_{j}\rangle\langle\tilde{\beta}_{j}|L^{\dagger}_1
+ |\xi_{j}|^2|\tilde{\alpha}_{j}\rangle\langle\tilde{\alpha}_{j}|  \right]
\end{eqnarray}
and $|\tilde{\alpha}_{j}\rangle=\frac{|{\alpha}_{j}\rangle}{\sqrt{\mathrm{Tr}\rho_{j}}}$,
$|\tilde{\beta}_{j}\rangle=\frac{|{\beta}_{j}\rangle}{\sqrt{\mathrm{Tr}\rho_{j}}}$.
Now, by making use of the property
\begin{equation}
\frac{1}{\mathrm{Tr}\rho_{j+1}} =\frac{1}{ \mathrm{Tr}\rho_j  }    \left(1+k_{j} \tau \right) + O(\tau^2) ,
\end{equation}
we obtain the difference equation for the normalized {\it a posteriori} density operator,
\begin{eqnarray}
\tilde{\rho}_{j+1} &=& \tilde{\rho}_{j} +  \tilde{\rho}_{j} k_{j}\tau-i[H_{\mathcal{S}},\tilde{\rho}_{j}]\tau-\frac{1}{2}\{L^{\dagger}_1L_1+L^{\dagger}_2L_2, \tilde{\rho}_{j}\}\tau\nonumber\\&&-|\tilde{\beta}_{j}\rangle\langle \tilde{\alpha}_{j}|L_1\xi_{j}^{\ast}\tau-L^{\dagger}_1|\tilde{\alpha}_{j}\rangle\langle\tilde{\beta}_{j}|\xi_{j}\tau
-|\tilde{\alpha}_{j}\rangle\langle\tilde{\alpha}_{j}||\xi_{j}|^2\tau.
\end{eqnarray}
In a similar way, one can find the following coupled difference equations
\begin{eqnarray}
|\tilde{\alpha}_{j+1}\rangle\langle\tilde{\beta}_{j+1}|
&=&|\tilde{\alpha}_{j}\rangle\langle\tilde{\beta}_{j}|+|\tilde{\alpha}_{j}\rangle\langle\tilde{\beta}_{j}|k_{j}\tau
-i[H_{\mathcal{S}},|\tilde{\alpha}_{j}\rangle\langle\tilde{\beta}_{j}|]\tau
\nonumber\\&&-\frac{1}{2}\{L^{\dagger}_1L_1+L^{\dagger}_2L_2,|\tilde{\alpha}_{j}\rangle\langle\tilde{\beta}_{j}|\}\tau-|\tilde{\alpha}_{j}\rangle\langle\tilde{\alpha}_{j}|L_1\xi_{j}^{\ast}\tau,
\end{eqnarray}
\begin{eqnarray}
|\tilde{\alpha}_{j+1}\rangle\langle\tilde{\alpha}_{j+1}|&=&|\tilde{\alpha}_{j}\rangle\langle\tilde{\alpha}_{j}|
+|\tilde{\alpha}_{j}\rangle\langle\tilde{\alpha}_{j}|k_{j}\tau
-i[H_{\mathcal{S}},|\tilde{\alpha}_{j}\rangle\langle\tilde{\alpha}_{j}|]\tau
\nonumber\\&&- \frac{1}{2}\{L^{\dagger}_1L_1+L^{\dagger}_2L_2,|\tilde{\alpha}_{j}\rangle\langle\tilde{\alpha}_{j}|\}\tau.
\end{eqnarray}

If at time $(j+1)\tau$ the result is $\eta_{j+1}=(1,0)$, which means a count on the right and no count on the left, then  
\begin{eqnarray}
\rho_{j+1}&=&L_1\rho_{j}L^{\dagger}_{1}\tau+L_1|\beta_{j}\rangle\langle\alpha_{j}|
\xi_{j}^{\ast}\tau+|\alpha_{j}\rangle\langle\beta_{j}|L^{\dagger}_1\xi_{j}\tau +
|\alpha_{j}\rangle\langle\alpha_{j}||\xi_{j}|^2\tau ,
\end{eqnarray}
\begin{equation}
|\alpha_{j+1}\rangle\langle\beta_{j+1}|
=L_1|\alpha_{j}\rangle\langle\beta_{j}|L^{\dagger}_1\tau+L_1|\alpha_{j}\rangle\langle\alpha_{j}|\xi_{j}^{\ast}\tau,
\end{equation}
\begin{equation}
|\alpha_{j+1}\rangle\langle\alpha_{j+1}|
=L_1|\alpha_{j}\rangle\langle\alpha_{j}|L^{\dagger}_1\tau.
\end{equation}
We neglected higher order terms in $\tau$ in the above formulae. Thus the conditional probability of the result $(1,0)$ at time $(j+1)\tau$ given that the {\it a posteriori} state of $\mathcal{S}$ at $j\tau$ is $\tilde{\rho}_{j}$ has the following form
\begin{equation}
p_{j+1}((1,0)|\tilde{\rho}_{j})=k_{1,j}\tau,
\end{equation}
where 
\begin{equation}
k_{1,j} = \mathrm{Tr}\left(L^{\dagger}_1L_1\tilde{\rho}_{j}
+ \xi_{j}^{\ast}L_1|\tilde{\beta}_{j}\rangle\langle\tilde{\alpha}_{j} |
+\xi_{j}
|\tilde{\alpha}_{j}\rangle\langle\tilde{\beta}_{j}|L^{\dagger}_1
+|\xi_{j}|^2\,|\tilde{\alpha}_{j}\rangle\langle\tilde{\alpha}_{j}|  \right).
\end{equation}
Therefore, for the normalized density operator, we get
\begin{equation}
\tilde{\rho}_{j+1}=
\frac{1}{k_{1,j}}\bigg(L_1\tilde{\rho}_{j}L^{\dagger}_1
+L_1|\tilde{\beta}_{j}\rangle\langle\tilde{\alpha}_{j}|\xi_{j}^{\ast}	
+|\tilde{\alpha}_{j}\rangle\langle\tilde{\beta}_{j}|L^{\dagger}_1\xi_{j}+|\tilde{\alpha}_{j}\rangle\langle\tilde{\alpha}_{j}||\xi_{j}|^2\bigg)
\end{equation}
together with
\begin{equation}
|\tilde{\alpha}_{j+1}\rangle\langle\tilde{\beta}_{j+1}|
=\frac{1}{k_{1,j}}\left(L_1|\tilde{\alpha}_{j}\rangle\langle\tilde{\beta}_{j}|L^{\dagger}_1+
L_1|\tilde{\alpha}_{j}\rangle\langle\tilde{\alpha}_{j}|\xi_{j}^{\ast}\right),
\end{equation}
\begin{equation}
|\tilde{\alpha}_{j+1}\rangle\langle\tilde{\alpha}_{j+1}|
=\frac{1}{k_{1,j}}L_1|\tilde{\alpha}_{j}\rangle\langle\tilde{\alpha}_{j}|L^{\dagger}_1.
\end{equation}
If $\eta_{j+1}=(0,1)$, then we obtain, by (\ref{rec3a}) and (\ref{rec3b}), the formulae
\begin{equation}
\rho_{j+1}=L_2\rho_{j}L^{\dagger}_{2}\tau,
\end{equation}
\begin{equation}
|\alpha_{j+1}\rangle\langle\beta_{j+1}|
=L_2|\alpha_{j}\rangle\langle\beta_{j}|L^{\dagger}_2\tau,
\end{equation}
\begin{equation}
|\alpha_{j+1}\rangle\langle\alpha_{j+1}|
=L_2|\alpha_{j}\rangle\langle\alpha_{j}|L^{\dagger}_2\tau.
\end{equation}
The result $(0,1)$ means a photon on the left and no photon on the right.  
Hence the conditional probability of the outcome $(0,1)$ at $(j+1)\tau$ given $\tilde{\rho}_{j}$ is 
\begin{equation}
p_{j+1}((0,1)|\tilde{\rho}_{j})=k_{2,j}\tau,
\end{equation}
where 
\begin{eqnarray}
k_{2,j}= \mathrm{Tr}\left(L^{\dagger}_2L_2\tilde{\rho}_{j}\right).
\end{eqnarray}

Finally, we have
\begin{equation}
\tilde{\rho}_{j+1}=
\frac{1}{k_{2,j}}L_2\tilde{\rho}_{j}L^{\dagger}_2
\end{equation}
\begin{equation}
|\tilde{\alpha}_{j+1}\rangle\langle\tilde{\beta}_{j+1}|
=\frac{1}{k_{2,j}}L_2|\tilde{\alpha}_{j}\rangle\langle\tilde{\beta}_{j}|L^{\dagger}_2,
\end{equation}
\begin{equation}
|\tilde{\alpha}_{j+1}\rangle\langle\tilde{\alpha}_{j+1}|
=\frac{1}{k_{2,j}}L_2|\tilde{\alpha}_{j}\rangle\langle\tilde{\alpha}_{j}|L^{\dagger}_2.
\end{equation}
Note that if $\mathcal{S}$ is a two-level system, then its conditional state immediately after a count on the left is simply the ground state. Retardation from the source to the detector is omitted here, but this could be easily incorporated into our model.  
 
Now by combing these results, we get the set of coupled discrete filtering equations of the form: 
	\begin{eqnarray}\label{I}
	\tilde{\rho}_{j+1}&=& \tilde{\rho}_{j}-i[H_{\mathcal{S}},\tilde{\rho}_{j}]\tau-\frac{1}{2}\{L^{\dagger}_1L_1+L^{\dagger}_2L_2,\tilde{\rho}_{j}\}\tau
	+L_1\tilde{\rho}_{j}L^{\dagger}_1\tau+L_2\tilde{\rho}_{j}L^{\dagger}_2\tau
	\nonumber\\&&+ [|\tilde{\alpha_{j}}\rangle\langle\tilde{\beta}_{j}|,L^{\dagger}_1]\xi_{j}\tau
	+[L_1, |\tilde{\beta}_{j}\rangle\langle\tilde{\alpha}_{j}|]\xi^{\ast}_{j}\tau \nonumber\\
	&&+\bigg\{\frac{1}{k_{1,j}}\left(
	L_1\tilde{\rho}_{j}L^{\dagger}_1+L_1|\tilde{\beta}_{j}\rangle\langle\tilde{\alpha}_{j}|\xi^{\ast}_{j}+|\tilde{\alpha}_{j}\rangle\langle\tilde{\beta}_{j}| L^{\dagger}_1\xi_{j}\right.\nonumber\\&&\left.+
	|\tilde{\alpha}_{j}\rangle\langle\tilde{\alpha}_{j}||\xi_{j}|^2\right)
	-\tilde{\rho}_{j}\bigg\}(\Delta n_{1,j}-k_{1,j}\tau)\nonumber\\
	&&+\bigg\{\frac{L_2\tilde{\rho}_{j}L^{\dagger}_2}{k_{2,j}}
	-\tilde{\rho}_{j}\bigg\}(\Delta n_{2,j}-k_{2,j}\tau)
	\end{eqnarray}	
	\begin{eqnarray}  \label{II}
	|\tilde{\alpha}_{j+1}\rangle\langle\tilde{\beta}_{j+1}| &=& |\tilde{\alpha}_{j}\rangle\langle\tilde{\beta}_{j}|
	-i[H_{\mathcal{S}},|\tilde{\alpha}_{j}\rangle\langle\tilde{\beta}_{j}|]\tau 
	-\frac{1}{2}\{L^{\dagger}_1L_1+L^{\dagger}_2L_2,|\tilde{\alpha}_{j}\rangle\langle\tilde{\beta}_{j}|\}\tau \nonumber\\
	&&+L_1|\tilde{\alpha}_{j}\rangle\langle\tilde{\beta}_{j}|L^{\dagger}_1\tau +L_2|\tilde{\alpha}_{j}\rangle\langle\tilde{\beta}_{j}|L^{\dagger}_2\tau+ [L_1,|\tilde{\alpha}_{j}\rangle\langle\tilde{\alpha}_{j}|]\xi^{\ast}_{j}\tau\nonumber\\
	&&+ \bigg\{\frac{1}{k_{1,j}}\left(L_1|\tilde{\alpha}_{j}\rangle\langle\tilde{\beta}_{j}|L^{\dagger}_1+
	L_1|\tilde{\alpha}_{j}\rangle\langle\tilde{\alpha}_{j}|
	\xi^{\ast}_{j}\right) \nonumber\\&&-|\tilde{\alpha}_{j}\rangle\langle\tilde{\beta}_{j}|\bigg\}(\Delta n_{1,j}-k_{1,j}\tau)\nonumber\\
	&&+ \bigg\{\frac{L_2|\tilde{\alpha}_{j}\rangle\langle\tilde{\beta}_{j}|L^{\dagger}_2}{k_{2,j}} -|\tilde{\alpha}_{j}\rangle\langle\tilde{\beta}_{j}|\bigg\}(\Delta n_{2,j}-k_{2,j}\tau),
	\end{eqnarray}	
	\begin{eqnarray}   \label{III}
	|\tilde{\alpha}_{j+1}\rangle\langle\tilde{\alpha}_{j+1}|&=&
	|\tilde{\alpha}_{j}\rangle\langle\tilde{\alpha}_{j}|-i\left[H_{\mathcal{S}},|\tilde{\alpha}_{j}\rangle\langle\tilde{\alpha}_{j}|\right]\tau
	-\frac{1}{2}\left\{L^{\dagger}_1L_1+L^{\dagger}_2L_2,|\tilde{\alpha}_{j}\rangle\langle\tilde{\alpha}_{j}|\right\}\tau\nonumber\\
	&&+L_1|\tilde{\alpha}_{j}\rangle\langle\tilde{\alpha}_{j}|L^{\dagger}_1\tau+L_2|\tilde{\alpha}_{j}\rangle\langle\tilde{\alpha}_{j}|L^{\dagger}_2\tau\nonumber\\
	&&+\bigg\{\frac{L_1|\tilde{\alpha}_{j}\rangle\langle\tilde{\alpha}_{j}|L^{\dagger}_1}{k_{1,j}}
	-|\tilde{\alpha}_{j}\rangle\langle\tilde{\alpha}_{j}|\bigg\}(\Delta n_{1,j}-k_{1,j}\tau)\nonumber\\
	&&+\bigg\{\frac{L_2|\tilde{\alpha}_{j}\rangle\langle\tilde{\alpha}_{j}|L^{\dagger}_2}{k_{2,j}}
	-|\tilde{\alpha}_{j}\rangle\langle\tilde{\alpha}_{j}|\bigg\}(\Delta n_{2,j}-k_{2,j}\tau)
	\end{eqnarray}
with the initial conditions: $\tilde{\rho}_{j=0}=|\psi_{0}\rangle\langle\psi_{0}|$, $|\tilde{\alpha}_{j=0}\rangle\langle\tilde{\beta}_{j=0}|=0$, and $|\tilde{\alpha}_{j=0}\rangle\langle\tilde{\alpha}_{j=0}|=|\psi_{0}\rangle\langle\psi_{0}|$. For the increments of the stochastic processes, defined as 
\begin{equation}
\Delta {n}_{1,j}=n_{1,j+1}-n_{1,j},\;\;\;\Delta {n}_{2,j}=n_{2,j+1}-n_{2,j},
\end{equation}
we obtain the conditional expectations
\begin{equation}
\mathbbm{E}[\Delta n_{1,j}|\tilde{\rho}_{j}]= k_{1,j}\tau + O(\tau^2),\;\;\;
\mathbbm{E}[\Delta n_{2,j}|\tilde{\rho}_{j}]= k_{2,j}\tau + O(\tau^2).
\end{equation}
Let us make it clear that for the case of $\Delta n_{1,j}=1$ and $\Delta n_{2,j}=0$ as well as  $\Delta n_{1,j}=0$ and $\Delta n_{2,j}=1$, all terms in (\ref{I})-(\ref{III}) containing the infinitesimal $\tau$ are negligible. 

In the continuous-time limit i.e. when $N\to \infty$ and $\tau\to 0$ such that $T=\tau N$ is fixed, we obtain from (\ref{sp}) the two-dimensional stochastic process
\begin{equation}\label{csp}
n_{t}=\left(n_{1,t},n_{2,t}\right)
\end{equation}  
describing the continuous in-time detection of photons at the right and left outputs. Finally, we take the limit of $T\to +\infty$ and obtain the continuous-time amplitude $\xi_{t}$ with the normalization condition
\begin{equation}
\int_{0}^{+\infty}dt|\xi_{t}|^2 =1.
\end{equation}

From (\ref{I})-(\ref{III}), we get in the continuous-time limit the set of the differential filtering equations
	\begin{eqnarray}\label{filcont1}
	d\tilde{\rho}_{t}&=& \mathcal{L}(\tilde{\rho}_{t})+[\tilde{\rho}^{01}_{t},L^{\dagger}_1]\xi_{t}dt
	+[L_1, \tilde{\rho}^{10}_{t}]\xi^{\ast}_{t}dt\nonumber\\
	&&+ \bigg\{\frac{1}{k_{1,t}}\left(
	L_1\tilde{\rho}_{t}L^{\dagger}_1+L_1\tilde{\rho}^{10}_{t}\xi^{\ast}_{t}+\tilde{\rho}^{01}_{t}L^{\dagger}_1\xi_{t}
	+\tilde{\rho}^{00}_{t}|\xi_{t}|^2\right)
	-\tilde{\rho}_{t}\bigg\}\left(dn_{1,t}-k_{1,t}dt\right)
	\nonumber\\
	&&+ \bigg\{\frac{L_2\tilde{\rho}_{t}L^{\dagger}_2}{k_{2,t}}
-\tilde{\rho}_{t}\bigg\}\left(dn_{2,t}-k_{2,t}dt\right),
	\end{eqnarray}
	\begin{eqnarray}\label{filcont2}
	d\tilde{\rho}^{01}_{t}&=& \mathcal{L}(\tilde{\rho}^{01}_{t})+ [L_1,\tilde{\rho}^{00}_{t}]\xi^{\ast}_{t}dt \nonumber\\
	&&+\left\{\frac{1}{k_{1,t}}\left(L_1\tilde{\rho}^{01}_{t}L^{\dagger}_1+
	L_1\tilde{\rho}^{00}_{t}\xi^{\ast}_{t}\right)-\tilde{\rho}^{01}_{t}\right\}\left(dn_{1,t}-k_{1,t}dt\right)\noindent \\
	&&+\left\{\frac{L_2\tilde{\rho}^{01}_{t}L^{\dagger}_2}{k_{2,t}}-\tilde{\rho}^{01}_{t}\right\}\left(dn_{2,t}-k_{2,t}dt\right)
	\end{eqnarray}
	\begin{eqnarray}\label{filcont3}
	d\tilde{\rho}^{10}_{t}&=& \mathcal{L}(\tilde{\rho}^{10}_{t}) +\left[\tilde{\rho}^{00}_{t},L_1^{\dagger}\right]\xi_{t}dt \nonumber\\
	&&+\left\{\frac{1}{k_{1,t}}\left(L_1\tilde{\rho}^{10}_{t}L^{\dagger}_1+
	\tilde{\rho}^{00}_{t}L_{1}^{\dagger}\xi_{t}\right)-\tilde{\rho}^{10}_{t}\right\}\left(dn_{1,t}-k_{1,t}dt\right)\noindent \\
	&&+\left\{\frac{L_2\tilde{\rho}^{10}_{t}L^{\dagger}_2}{k_{2,t}}-\tilde{\rho}^{10}_{t}\right\}\left(dn_{2,t}-k_{2,t}dt\right)
	\end{eqnarray}
		\begin{eqnarray}\label{filcont4}
	d\tilde{\rho}^{00}_{t}&=&\mathcal{L}(\tilde{\rho}^{00}_{t})+ \left\{\frac{L_1\tilde{\rho}^{00}_{t}L^{\dagger}_1}{k_{1,t}}-
	\tilde{\rho}^{00}_{t}\right\}\left(dn_{1,t}-k_{1,t}dt\right)\nonumber\\
	&&+ \left\{\frac{L_2\tilde{\rho}^{00}_{t}L^{\dagger}_2}{k_{2,t}}-
	\tilde{\rho}^{00}_{t}\right\}\left(dn_{2,t}-k_{2,t}dt\right)
	\end{eqnarray}
	with the superoperator 
	\begin{equation}
	\mathcal{L}\rho=-i[H_{\mathcal{S}},\rho]-
	\frac{1}{2}\{L^{\dagger}_1L_1+L^{\dagger}_2L_2,{\rho}\}		+L_1{\rho}L^{\dagger}_1+L_2{\rho}L^{\dagger}_2
	\end{equation}
and initially  $\tilde{\rho}_{t=0}=|\psi_{0}\rangle\langle\psi_{0}| $, $\tilde{\rho}^{01}_{t=0}= 0$, $\tilde{\rho}^{10}_{t=0}= 0$, and $\tilde{\rho}^{00}_{t=0}=|\psi_{0}\rangle\langle\psi_{0}|$. Let us stress that the infinitesimal time interval is short enough that the probability of detecting two or more photons is negligible, see also \cite{Srinivas81,Mollow68,Car93,WM10,Baragiola17}. Thus the It\^{o} rule for $dn_{l,t}=n_{l,t+td}-n_{l,t}$ in this case is
\begin{equation}
dn_{l,t}dn_{i,t}=\delta_{li}dn_{l,t}. 
\end{equation}
Note that when one of the $dn_{l,t}$ is equal to one then all the other terms in the r.h.s. of (\ref{filcont1})-(\ref{filcont4}) are negligible.
For the increment of the stochastic processes, we have the conditional means
\begin{equation}\label{meancon1}
\mathbbm{E}[dn_{1,t}|\tilde{\rho}_{t}]=k_{1,t}dt,
\end{equation}
\begin{equation}\label{meancon2}
\mathbbm{E}[dn_{2,t}|\tilde{\rho}_{t}]=k_{2,t}dt,
\end{equation}
where
\begin{equation}\label{inten1}
k_{1,t}=\mathrm{Tr}\left(L^{\dagger}_1L_1\tilde{\rho}_{t}
+L_1\tilde{\rho}^{10}_{t}\xi_{t}^{\ast}+\tilde{\rho}^{01}_{t}L^{\dagger}_1\xi_{t}+\tilde{\rho}^{00}_{t}|\xi_{t}|^2\right),
\end{equation}
\begin{equation}\label{inten2}
k_{2,t}=\mathrm{Tr}\left(L^{\dagger}_2L_2\tilde{\rho}_{t}\right).
\end{equation}
 The expressions (\ref{meancon1}) and (\ref{meancon2}) are respectively the probabilities of detecting a photon at the right and left detector in the time interval $[t,t+dt)$. They are the conditional probabilities depending on all results of the past measurements up to $t$. Clearly, they also define the conditional mean number of photons detected on the right and the left in $[t,t+dt)$, respectively. In (\ref{inten1}) and (\ref{inten2}) one can easily recognize the sources of photons in the output fields. In (\ref{inten1}) the first term is associated with the photons emitted  by the quantum system to the right, the last term is the free  input field, and the remaining terms appear due to the interference between the two. While in (\ref{inten2}), we deal with the mean intensity of the field emitted by $\mathcal{S}$ to the left.

By taking an average over all trajectories, we obtain from (\ref{filcont1})-(\ref{filcont4}) the set of four coupled differential equations describing the {\it a priori} evolution of $\mathcal{S}$:
\begin{equation}\label{apriori1}
\dot{\varrho}_{t}=\mathcal{L}(\varrho_{t})
+[\varrho^{01}_{t},L^{\dagger}_1]\xi_{t}
+[L_1, {\varrho}^{10}_{t}]\xi^{\ast}_{t},
\end{equation}
\begin{equation}
\dot{\varrho}^{01}_{t}=\mathcal{L}(\varrho^{01}_{t})+ \left[L_1,{\varrho}^{00}_{t}\right]\xi^{\ast}_{t},
\end{equation}
\begin{equation}
\dot{\varrho}^{10}_{t}=\mathcal{L}(\varrho^{10}_{t})+ \left[{\varrho}^{00}_{t},L_1^{\dagger}\right]\xi_{t},
\end{equation}
\begin{equation}\label{apriori2}
\dot{\varrho}^{00}_{t}= \mathcal{L}(\varrho^{00}_{t})
\end{equation}
with $\varrho_{t=0}=|\psi_{0}\rangle\langle\psi_{0}|$, $\varrho^{01}_{t=0}= 0$, $\varrho^{10}_{t=0}= 0$, and $\varrho^{00}_{t=0}=|\psi_{0}\rangle\langle\psi_{0}|$. Of course, the set of the filtering equations, as well as the set of master equations, are valid for an arbitrary initial state of $\mathcal{S}$.


Note that if $L_{2}=0$ then one obtains from (\ref{filcont1})-(\ref{filcont4}) and (\ref{apriori1})-(\ref{apriori2}), respectively, the conditional and unconditional evolution of $\mathcal{S}$ for a unidirectional field prepared in a single-photon state, and both mentioned sets reduce to the known sets of equations derived in the papers \cite{GEPZ98,Baragiola2012,Gough12a,Gough12b,Gough13,Baragiola17,Dabrowska17,Dabrowska18}. Moreover, Eqs. (\ref{apriori1})-(\ref{apriori2}) are consistent with the set of two-mode Fock-state master equations, determined by making use of quantum stochastic It\^{o} calculus, in \cite{Baragiola2012}.    
If $\xi(t)=0$, i.e. there is no input photon, then (\ref{filcont1})-(\ref{filcont4}) reduces to a single filtering and (\ref{apriori1})-(\ref{apriori2}) to a single master equation for a quantum system coupled to two vacuum inputs. 

\section{Conditional vectors and statistics of counts for continuous-time observation}


To characterize the stochastic counting process (\ref{csp}) by the exclusive probability densities, we determine the general analytical formulae for the conditional vectors associated with different realisations of (\ref{csp}). The formulae for the conditional vectors defining quantum trajectories for the continuous in time observation of the output field, we obtain from the discrete vectors given in \ref{condivectors}. One can check that continuous-time versions of the conditional vectors for some chosen trajectories have the following form: 
\begin{enumerate}
	\item for zero counts from $0$ to $t$ one gets from (\ref{convec1a}) and (\ref{convec1b}) respectively:
	\begin{equation}\label{convec8a}
	|\alpha_{t|0}\rangle = \mathbf{T}_t |\psi_{0}\rangle
	\end{equation}
	and
	\begin{eqnarray}\label{convec8b}
	|\beta_{t|0}\rangle = -\int_{0}^{t}ds \mathbf{T}_{t-s}\xi_{s}L^{\dagger}_{1} \mathbf{T}_{s}|\psi_{0}\rangle
	\end{eqnarray}
	where
	\begin{equation}\label{}
	\mathbf{T}_t = e^{-iGt} ,
	\end{equation}
	is a non-unitary propagator with a non-hermitian Hamiltonian $G=H_{\mathcal{S}}-\frac{i}{2}\left(L^{\dagger}_{1}L_{1}+L^{\dagger}_{2}L_{2} \right)$,
	\item for one count at the time $t^{\prime}$ at the right detector and no other counts in the interval $(0,t]$ one gets from (\ref{convec2a}) and (\ref{convec2b}) respectively:
	\begin{equation}\label{convec9a}
	|\alpha_{t|R,t^{\prime}}\rangle = \sqrt{dt^{\prime}}\mathbf{T}_{t-t^{\prime}} L_{1} \mathbf{T}_{t^{\prime}} |\psi_{0}\rangle ,
	\end{equation}
	\begin{eqnarray}\label{convec9b}
	|\beta_{j|R,t^{\prime}}\rangle &=&  \sqrt{dt^{\prime}}\Big[  \mathbf{T}_t \xi_{t^{\prime}} - \mathbf{T}_{t-t^{\prime}} L_{1} \int_{0}^{t^{\prime}}ds \, \mathbf{T}_{t^{\prime}-s} \xi_{s} L^{\dagger}_{1} \mathbf{T}_s \nonumber\\&&-\int_{t^{\prime}}^{t} ds \mathbf{T}_{t-s} \xi_{s} L^{\dagger}_{1} \mathbf{T}_{s-t^{\prime}} L_{1} \mathbf{T}_{t^{\prime}}\Big]|\psi_{0}\rangle,
	\end{eqnarray}
	\item for two counts at  $t^{\prime}$ and $t^{\prime\prime}$ such that $0<t^{\prime}<t^{\prime\prime}$ both at the right detector, and no other counts in the interval $(0,t]$, one finds from (\ref{convec3a}) and (\ref{convec3b}) respectively:
	\begin{equation}\label{convec10a}
	|\alpha_{t|R,t^{\prime\prime};R, t^{\prime}}\rangle=\sqrt{dt^{\prime\prime}dt^{\prime}}
	\mathbf{T}_{t-t^{\prime\prime}} L_{1} \mathbf{T}_{t^{\prime\prime}-t^{\prime}} L_{1} \mathbf{T}_{t^{\prime}}|\psi_{0}\rangle,
	\end{equation}
	\begin{eqnarray}\label{convec10b}
	|\beta_{t|R,t^{\prime\prime}; R,t^{\prime}}\rangle &=&\sqrt{dt^{\prime\prime}dt^{\prime}}
	\Big[
	\mathbf{T}_{t-t^{\prime\prime}} L_{1} \mathbf{T}_{t^{\prime\prime}} \xi_{t^{\prime}} + \mathbf{T}_{t-t^{\prime}} \xi_{t^{\prime\prime}}L_{1} \mathbf{T}_{t^{\prime}} \nonumber\\&&- \mathbf{T}_{t-t^{\prime\prime}} L_{1} \mathbf{T}_{t^{\prime\prime}-t^{\prime}} L_{1} \Big( \int_{0}^{t^{\prime}} ds \mathbf{T}_{t^{\prime}-s} \xi_{s} L^{\dagger}_{1} \mathbf{T}_s\Big) \\
	&&- \mathbf{T}_{t-t^{\prime\prime}} L_{1} \Big( \int_{t^{\prime}}^{t^{\prime\prime}}ds
	\mathbf{T}_{t^{\prime\prime}-s} \xi_{s} L^{\dagger}_{1} \mathbf{T}_{s-t^{\prime}} \Big) L_{1} \mathbf{T}_{t^{\prime}}
	\nonumber\\&&- \Big( \int_{t^{\prime\prime}}^{t}ds \mathbf{T}_{t-s} \xi_{s} L^{\dagger}_{1} \mathbf{T}_{s-t^{\prime\prime}} \Big)
	L_{1} \mathbf{T}_{t^{\prime\prime}-t^{\prime}} L_{1} \mathbf{T}_{t^{\prime}} \Big]|\psi_{0}\rangle,\nonumber
	\end{eqnarray}
	\item for one count at the time $t^{\prime}$ at the left detector and no other counts in the interval $(0,t]$ one obtains from (\ref{convec4a}) and (\ref{convec4b}) respectively: 
	\begin{equation}\label{convec11a}
	|\alpha_{t|L,t^{\prime}}\rangle = \sqrt{dt^{\prime}}\mathbf{T}_{t-t^{\prime}} L_{2} \mathbf{T}_{t^{\prime}} |\psi_{0}\rangle ,
	\end{equation}
	\begin{eqnarray}\label{convec11b}
	|\beta_{j|L,t^{\prime}}\rangle &=& \sqrt{dt^{\prime}} \Big[ - \mathbf{T}_{t-t^{\prime}} L_{2} \int_{0}^{t^{\prime}}ds \, \mathbf{T}_{t^{\prime}-s} \xi_{s} L^{\dagger}_{1} \mathbf{T}_s \nonumber\\&&- \int_{t^{\prime}}^{t} ds \mathbf{T}_{t-s} \xi_{s} L^{\dagger}_{1} \mathbf{T}_{s-t^{\prime}} L_{2} \mathbf{T}_{t^{\prime}}\Big]|\psi_{0}\rangle,
	\end{eqnarray}
	\item for two counts at  $t^{\prime}$ and $t^{\prime\prime}$ such that
	$0<t^{\prime}<t^{\prime\prime}$ both at the left detector, and no other counts in the interval $(0,t]$, one finds from (\ref{convec5a}) and (\ref{convec5b}) respectively:
	\begin{equation}\label{convec12a}
	|\alpha_{t|L,t^{\prime\prime}; L,t^{\prime}}\rangle=\sqrt{dt^{\prime\prime}dt^{\prime}}
	\mathbf{T}_{t-t^{\prime\prime}} L_{2} \mathbf{T}_{t^{\prime\prime}-t^{\prime}} L_{2} \mathbf{T}_{t^{\prime}}|\psi_{0}\rangle,
	\end{equation}
	\begin{eqnarray}\label{convec12b}
	|\beta_{t|L,t^{\prime\prime};L, t^{\prime}}\rangle &=&\sqrt{dt^{\prime\prime}dt^{\prime}}
	\Big[-\mathbf{T}_{t-t^{\prime\prime}} L_{2}
\mathbf{T}_{t^{\prime\prime}-t^{\prime}} L_{2} \int_{0}^{t^{\prime}} ds \mathbf{T}_{t^{\prime}-s} \xi_{s} L^{\dagger}_{1} \mathbf{T}_s \nonumber \\
	&&- \mathbf{T}_{t-t^{\prime\prime}} L_{2} \Big( \int_{t^{\prime}}^{t^{\prime\prime}}ds
	\mathbf{T}_{t^{\prime\prime}-s} \xi_{s} L^{\dagger}_{1} \mathbf{T}_{s-t^{\prime}} \Big) L_{2} \mathbf{T}_{t^{\prime}}
	\nonumber\\&&- \Big( \int_{t^{\prime\prime}}^{t}ds \mathbf{T}_{t-s} \xi_{s} L^{\dagger}_{1} \mathbf{T}_{s-t^{\prime\prime}} \Big)
	L_{2} \mathbf{T}_{t^{\prime\prime}-t^{\prime}} L_{2} \mathbf{T}_{t^{\prime}} \Big]|\psi_{0}\rangle, \nonumber
	\end{eqnarray}
	\item for two counts at $t^{\prime}$ and $t^{\prime\prime}$ such that
	$0<t^{\prime}<t^{\prime\prime}$ taking place sequentially at the right and the left detector, and no other counts in the interval $(0,t]$, one finds from (\ref{convec6a}) and (\ref{convec6b}):
	\begin{equation}\label{convec13a}
	|\alpha_{t|L,t^{\prime\prime}; R,t^{\prime}}\rangle=\sqrt{dt^{\prime\prime}dt^{\prime}}
	\mathbf{T}_{t-t^{\prime\prime}} L_{2} \mathbf{T}_{t^{\prime\prime}-t^{\prime}} L_{1} \mathbf{T}_{t^{\prime}}|\psi_{0}\rangle,
	\end{equation}
	\begin{eqnarray}\label{convec13b}
	|\beta_{t|L,t^{\prime\prime};R, t^{\prime}}\rangle &=&\sqrt{dt^{\prime\prime}dt^{\prime}}
	\Big[\mathbf{T}_{t-t^{\prime\prime}} L_{2} \mathbf{T}_{t^{\prime\prime}-t^{\prime}} \xi_{t^{\prime}}\mathbf{T}_{t^{\prime}}
	\nonumber\\&&-\mathbf{T}_{t-t^{\prime\prime}} L_{2} \mathbf{T}_{t^{\prime\prime}-t^{\prime}} L_{1} \Big( \int_{0}^{t^{\prime}} ds \mathbf{T}_{t^{\prime}-s} \xi_{s} L^{\dagger}_{1} \mathbf{T}_s\Big)  \\
	&&- \mathbf{T}_{t-t^{\prime\prime}} L_{2} \Big( \int_{t^{\prime}}^{t^{\prime\prime}}ds
	\mathbf{T}_{t^{\prime\prime}-s} \xi_{s} L^{\dagger}_{1} \mathbf{T}_{s-t^{\prime}} \Big) L_{1} \mathbf{T}_{t^{\prime}}\nonumber\\
	&&- \Big( \int_{t^{\prime\prime}}^{t}ds \mathbf{T}_{t-s} \xi_{s} L^{\dagger}_{1} \mathbf{T}_{s-t^{\prime\prime}} \Big)
	L_{2} \mathbf{T}_{t^{\prime\prime}-t^{\prime}} L_{1} \mathbf{T}_{t^{\prime}} \Big]|\psi_{0}\rangle, \nonumber
	\end{eqnarray}
	\item for two counts at  $t^{\prime}$ and $t^{\prime\prime}$ such that
	$0<t^{\prime}<t^{\prime\prime}$ taking place sequentially at the left and the right detector, and no other counts in the interval $(0,t]$, one finds from (\ref{convec7a}) and (\ref{convec7b}):
	\begin{equation}\label{convec14a}
	|\alpha_{t|R,t^{\prime\prime}; L,t^{\prime}}\rangle=\sqrt{dt^{\prime\prime}dt^{\prime}}
	\mathbf{T}_{t-t^{\prime\prime}} L_{1} \mathbf{T}_{t^{\prime\prime}-t^{\prime}} L_{2} \mathbf{T}_{t^{\prime}}|\psi_{0}\rangle,
	\end{equation}
	\begin{eqnarray}\label{convec14b}
|\beta_{t|R,t^{\prime\prime}; L,t^{\prime}}\rangle &=&\sqrt{dt^{\prime\prime}dt^{\prime}}
	\Big[\mathbf{T}_{t-t^{\prime}} \xi_{t^{\prime\prime}} L_{2} \mathbf{T}_{t^{\prime}}
	\nonumber\\&&-\mathbf{T}_{t-t^{\prime\prime}} L_{1} \mathbf{T}_{t^{\prime\prime}-t^{\prime}} L_{2} \Big( \int_{0}^{t^{\prime}} ds \mathbf{T}_{t^{\prime}-s} \xi_{s} L^{\dagger}_{1} \mathbf{T}_s\Big)  \\
	&&- \mathbf{T}_{t-t^{\prime\prime}} L_{1} \Big( \int_{t^{\prime}}^{t^{\prime\prime}}ds
	\mathbf{T}_{t^{\prime\prime}-s} \xi_{s} L^{\dagger}_{1} \mathbf{T}_{s-t^{\prime}} \Big) L_{2} \mathbf{T}_{t^{\prime}}\nonumber\\&&- \Big( \int_{t^{\prime\prime}}^{t}ds \mathbf{T}_{t-s} \xi_{s} L^{\dagger}_{1} \mathbf{T}_{s-t^{\prime\prime}} \Big)
	L_{1} \mathbf{T}_{t^{\prime\prime}-t^{\prime}} L_{2} \mathbf{T}_{t^{\prime}} \Big]|\psi_{0}\rangle. \nonumber
	\end{eqnarray}
	
\end{enumerate}
One can easily provide a physical interpretation of the above expressions. 
Of course, one has to also take care of the operators ordering. The propagator $\mathbf{T}_{t}$ refers to the period of time when there are no photon emissions and no photon absorption. The term $-\xi_{t}L_{1}^{\dagger}$ describes the process of absorption of the input photon by $\mathcal{S}$.  Intervals without photon emissions and absorption are interrupted by some jumps. The operators $L_{1}$ and $L_{2}$ are associated with the processes of emissions of photons by $\mathcal{S}$ respectively to the left and to the right. The quantity $\xi_{t}$ appearing between two propagators is associated with direct detection of the photon of the input field.  It is difficult to write down general formulae for the conditional vectors, but their structure is easy to recognize and describe.

The {\it a priori} state of $\mathcal{S}$ (solution to Eqs. (\ref{apriori1})-(\ref{apriori2})) in the representation of the counting stochastic process (\ref{csp}) has the form 
\begin{equation}\label{aprior}
\varrho_{t}=\rho_{t|0}+\sum_{m=1}^{+\infty}\int_{0}^{t}\!dt_{m}
\!\int_{0}^{t_{m}}\!dt_{m-1}\!\ldots\!
\int_{0}^{t_{2}}\!dt_{1}
\!\sum_{D_{m},D_{m-1},\ldots,D_{1}=R,L}\!\!\!\!\!\!\!\rho_{t|{D_{m}},t_{m};D_{m-1},t_{m-1};\ldots;{D_1},t_{1}}
\end{equation}
with the conditional unnormalised operators defined by 
\begin{equation}
\rho_{t|0}=|\alpha_{t|0}\rangle\langle\alpha_{t|0}|\int_{t}^{+\infty}ds|\xi_{s}|^2+|\beta_{t|0}\rangle\langle\beta_{t|0}|,
\end{equation}
and	
\begin{eqnarray}
\lefteqn{dt_{m}dt_{m-1}\ldots dt_{1}\rho_{t|{D_{m}},t_{m};D_{m-1},t_{m-1};\ldots;{D_1},t_{1}}}
\nonumber\\&&\;\;=|\alpha_{t|{D_{m}},t_{m};D_{m-1},t_{m-1};\ldots;{D_1},t_{1}}\rangle\langle\alpha_{t|{D_{m}},t_{m};D_{m-1},t_{m-1};\ldots;{D_1},t_{1}}|\int_{t}^{+\infty}ds|\xi_{s}|^2
\nonumber\\ &&\;\;+|\beta_{t|{D_{m}},t_{m};D_{m-1},t_{m-1};\ldots;{D_1},t_{1}}\rangle\langle\beta_{t|{D_{m}},t_{m};D_{m-1},t_{m-1};\ldots;{D_1},t_{1}}|.
\end{eqnarray}
Note that in (\ref{aprior}) we have a sum over all the photon detection pathways taking place in the interval from $0$ to $t$. The quantity
\begin{eqnarray}\label{zerocounts}
P_{0}^{t}(0)=\mathrm{Tr} \rho_{t|0}
\end{eqnarray}
is the probability of having no counts from $0$ up to $t$ and
\begin{eqnarray}\label{condensity}
p_{0}^{t}\left({D_{m}},t_{m};{D_{m-1}},t_{m-1};\ldots;{D_1},t_{1}\right)=
\mathrm{Tr}\rho_{t|{D_{m}},t_{m};D_{m-1},t_{m-1};\ldots;{D_1},t_{1}}
\end{eqnarray}
defines the exclusive probability density of $m$ counts each in one of the nonoverlapping intervals $[t_{1}+dt_1), [t_{2}+dt_2),\ldots, [t_{m}+dt_{m})$, such that  $0<t_{1}< t_{2}<\ldots< t_{m}\leq t$, taking place respectively at the detectors  $D_{1}, D_{2}, \ldots, D_{m}$, and no other counts in the interval $(0,t]$. Hence, we obtain the counting formula  
	\begin{eqnarray}
P_{0}^{t}(m;D_{m},\ldots,D_2,D_{1})=\int_{0}^{t}dt_{m}\int_{0}^{t_{m}}dt_{m-1}\ldots
	\int_{0}^{t_{2}}dt_{1}
	 p_{0}^{t}\left({D_{m}},t_{m};{D_{m-1}},t_{m-1};\ldots;{D_1},t_{1}\right).
	\end{eqnarray}
for the probability of $m$ counts in the interval $(0,t]$ registered at the detectors $D_{1}$, $D_2$, $D_{m}$. 

The exclusive probability density allows us to construct the whole statistics of the counts.
Using the quantum trajectories, one can determine such quantities as conditional and unconditional mean intensities of the field and waiting-time distribution for photons in the output field.  

\section{Example: The output field for a two-level system}

Let us consider the system $\mathcal{S}$ that is a two-level atom. By $|g\rangle$ and $|e\rangle$ we denote respectively the ground and the excited states of the system. We define its interaction with a bidirectional electromagnetic field by the coupling operators
\begin{equation}\label{L1}
L_{1}=\sqrt{\Gamma_{1}}\sigma_{-},
\end{equation} 
\begin{equation}\label{L2}
L_{2}=\sqrt{\Gamma_{2}}\sigma_{-},
\end{equation} 
where $\sigma_{-}=|g\rangle \langle e|$ and $\Gamma_1,\Gamma_2$ are non-negative coupling constants. The Hamiltonian of the system, written in the rotating frame, has the form
\begin{equation}\label{atom2}
H_{\mathcal{S}}=-\frac{\Delta_{0}}{2}\sigma_{z},
\end{equation}
where $\sigma_{z}=|e\rangle \langle e|-|g\rangle \langle g|$, and $\Delta_{0}=\omega_{c}-\omega_{0}$, where $\omega_{0}$ is the transition frequency of the atom and $\omega_{c}$ represents the carrier frequency of the input wave packet. 


The analytical formulae for the {\it a priori} state of the two-level atom for any initial state of the system, an arbitrary photon profile, and a unidirectional field was given in \cite{Dabrowska2020a}. The general solution to Eq. (\ref{apriori1})--(\ref{apriori2}) was given in \cite{Dabrowska2020}. We will not discuss these solutions here but it is clear that the atom is driven by the single-photon field and it asymptomatically relaxes to the ground state.  

\subsection{Statistics of counts in the output fields}

One can easily give an intuitive physical interpretation of the conditional vectors for the two-level atom. For example, from (\ref{convec8a}) and (\ref{convec8b}) it follows that if the atom was initially in the excited state and we did not observe any count up to $t$ it means that the atom has not met the qubit prepared in the excited state yet (the photon appears in the future) and the atom has stayed in the excited state up to $t$. If the atom is initially in the ground state it is possible that we do not observe any count up to $t$ because the atom has not met the input photon yet or it has already met this photon, absorbed it, and stayed in the excited state up to $t$. If we observed a photon on the right side at time $t^{\prime}$ and no other photons from $0$ to $t$ we deal with the following possibilities:  
\begin{itemize}
	\item the atom has not met the input photon yet and we observed a photon emitted by the atom (see (\ref{convec9a})),
	\item we observed directly the photon coming from the left or the atom had absorbed the input photon before $t^{\prime}$ and then emitted it at $t^{\prime}$, or the atom emitted a photon at $t^{\prime}$, then absorbed the input photon and has stayed in the excited state up to $t$ (see (\ref{convec9b})).  
\end{itemize} 
In a similar way one can easily characterize all the other conditional vectors defining quantum trajectories. 

Now by making use of (\ref{convec8a}), (\ref{convec8b}), and (\ref{zerocounts}) we derive the formula for the probability of not detecting any photon up to $t$:
\begin{eqnarray}\label{Pzero}
P_{t}(0)&=&e^{-\Gamma t}\int_{t}^{+\infty}ds |\xi_{s}|^2\rho_{ee}(0)\nonumber\\&&+\left(\int_{t}^{+\infty}ds |\xi_{s}|^2+\Gamma_{1} e^{-\Gamma t} \left|\int_{0}^{t}ds\xi_{s} e^{\left(-i\Delta_{0}+\frac{\Gamma}{2} \right)s}\right|^2\right)\varrho_{gg}(0),
\end{eqnarray}
where $\Gamma=\Gamma_1+\Gamma_2$ and
\begin{equation}
\varrho(0)  = \left( \begin{array}{cc} \varrho_{gg}(0)&  \varrho_{ge}(0) \\
 \varrho_{eg}(0)  & \varrho_{ee}(0) 
\end{array} \right).
\end{equation}
is an arbitrary initial state of the atom. One can easily check the following properties: $P_{t=0}(0)=1$ and $\lim_{t\to+\infty}P_{t}(0)=0$. By referring to (\ref{condensity}) one can find the probability density of detecting a photon at $t^{\prime}$ and no other photons from $0$ up to $t$: 
\begin{equation}
p_{0}^{t}(t^{\prime})=p_{0}^{t}\left(R,t^{\prime}\right)+p_{0}^{t}\left(L,t^{\prime}\right),
\end{equation}
where 
\begin{eqnarray}\label{pR}
&&p_{0}^{t}\left(R,t^{\prime}\right)= \left|\xi_{t^{\prime}}-\Gamma_{1}\int_{0}^{t^{\prime}}ds e^{\left(-i\Delta_{0} +\frac{\Gamma}{2}\right)(s-t^{\prime})}\xi_{s}\right|^2\rho_{gg}(0)\nonumber\\
&&+\bigg[\Gamma_{1}e^{-\Gamma t^{\prime}}\int_{t}^{+\infty}ds|\xi_{s}|^2 +e^{-\Gamma t}\left|\xi_{t^{\prime}}-\Gamma_{1}\int_{t^{\prime}}^{t}ds e^{\left(-i\Delta_{0} +\frac{\Gamma}{2}\right)(s-t^{\prime})}\xi_{s}\right|^2\bigg]\varrho_{ee}(0)\nonumber
\end{eqnarray}
follows from (\ref{convec9a}) and (\ref{convec9b}), and 
\begin{eqnarray}\label{pL}
&&p_{0}^{t}\left(L,t^{\prime}\right)= \Gamma_{1}\Gamma_{2}e^{-\Gamma t^{\prime}}\left|\int_{0}^{t^{\prime}}ds e^{\left(-i\Delta_{0} +\frac{\Gamma}{2}\right)s}\xi_{s}\right|^2\rho_{gg}(0)\nonumber\\
&&+\Gamma_{2}e^{-\Gamma t^{\prime}}\bigg[\int_{t}^{+\infty}ds|\xi_{s}|^2+\Gamma_{1}e^{-\Gamma t}\left|\int_{t^{\prime}}^{t}ds e^{\left(-i\Delta_{0} +\frac{\Gamma}{2}\right)s}\xi_{s}\right|^2\bigg]\varrho_{ee}(0)
\end{eqnarray}
from (\ref{convec11a}) and (\ref{convec11b}). By $p_{0}^{t}\left(R,t^{\prime}\right)$ we denoted the probability density of a count at $t^{\prime}$ on the right side and no other counts (on the left and on the right side) from $0$ to $t$. One can easily check the following intuitive properties --- namely, if initially the atom is in the excited state then $\lim_{t\to +\infty}p_{0}^{t}\left(R,t^{\prime}\right)=0$ and $\lim_{t\to +\infty}p_{0}^{t}\left(L,t^{\prime}\right)=0$, which means that we are certain  that in this case  more than one photon appears ultimately in the output fields.

The formula for the probability density of detecting two photons, respectively, at $t^{\prime}$ and $t^{\prime\prime}$ such that $0<t^{\prime}<t^{\prime\prime}\leq t$ and no other photons up to $t$ one can finds by
\begin{eqnarray}
p_{0}^{t}(t^{\prime\prime},t^{\prime})&=&
p_{0}^{t}\left(R,t^{\prime\prime};R,t^{\prime}\right)+
p_{0}^{t}\left(R,t^{\prime\prime};L,t^{\prime}\right)
+p_{0}^{t}\left(L,t^{\prime\prime};R,t^{\prime}\right)\nonumber\\&&+
p_{0}^{t}\left(L,t^{\prime\prime};L,t^{\prime}\right),
\end{eqnarray}
where 
\begin{eqnarray}
p_{0}^{t}\left(R,t^{\prime\prime};R,t^{\prime}\right)&=&\Gamma_{1}e^{-\Gamma(t^{\prime\prime}+t^{\prime})}\bigg| \xi_{t^{\prime}}e^{\left(-i\Delta_{0}+\frac{\Gamma}{2}\right)t^{\prime}}+\xi_{t^{\prime\prime}}e^{\left(-i\Delta_{0}+\frac{\Gamma}{2}\right)t^{\prime\prime}}
\nonumber\\&&-\Gamma_{1} 
	\int_{{t}^{\prime}}^{{t}^{\prime\prime}}ds  e^{\left(-i\Delta_{0}+\frac{\Gamma}{2}\right)s}\xi_{s}
	\bigg|^2\!\rho_{ee}(0)
\end{eqnarray}
follows from (\ref{convec10a}) and (\ref{convec10b}),
\begin{eqnarray}
&&p_{0}^{t}\left(L,t^{\prime\prime};R,t^{\prime}\right)
=\Gamma_{2}e^{-\Gamma t^{\prime\prime}}\bigg| \xi_{t^{\prime}}-\Gamma_{1} 
\int_{{t}^{\prime}}^{{t}^{\prime\prime}}ds  e^{\left(-i\Delta_{0}+\frac{\Gamma}{2}\right)(s-t^{\prime})}\xi_{s}
\bigg|^2\varrho_{ee}(0)
\end{eqnarray}
from (\ref{convec13a}) and (\ref{convec13b}),
\begin{eqnarray}
&&p_{0}^{t}\left(R,t^{\prime\prime};L,t^{\prime}\right)
=\Gamma_{2}e^{-\Gamma t^{\prime}}\left| \xi_{t^{\prime\prime}}-\Gamma_{1} 
\int_{{t}^{\prime}}^{{t}^{\prime\prime}}ds  e^{\left(-i\Delta_{0}+\frac{\Gamma}{2}\right)(s-t^{\prime\prime})}\xi_{s}
\right|^2\varrho_{ee}(0)
\end{eqnarray}
from (\ref{convec14a}) and (\ref{convec14b}), and
\begin{eqnarray}
&&p_{0}^{t}\left(L,t^{\prime\prime};L,t^{\prime}\right)
=\Gamma_{2}^2\Gamma_{1}
e^{-\Gamma(t^{\prime\prime}+t^{\prime})}\left| 
\int_{{t}^{\prime}}^{{t}^{\prime\prime}}ds  e^{\left(-i\Delta_{0}+\frac{\Gamma}{2}\right)s}\xi_{s}
\right|^2\varrho_{ee}(0)
\end{eqnarray}
from (\ref{convec12a}) and (\ref{convec12b}). Hence one can determine the formula
\begin{eqnarray}
p_{0}^{t}\left(t^{\prime\prime},t^{\prime}\right)&=&\bigg\{
\Gamma_{1}
e^{-\Gamma(t^{\prime\prime}+t^{\prime})}\bigg| \xi_{t^{\prime}}e^{-\left(i\Delta_{0}-\frac{\Gamma}{2}\right)t^{\prime}}+\xi_{t^{\prime\prime}}e^{-\left(i\Delta_{0}-\frac{\Gamma}{2}\right)t^{\prime\prime}}
\nonumber\\&&-\Gamma 
\int_{t^{\prime}}^{t^{\prime\prime}}ds  e^{\left(-i\Delta_{0}+\frac{\Gamma}{2}\right)s}\xi_{s}
\bigg|^2\nonumber\\
&&+\Gamma_{2}\left(e^{-\Gamma t^{\prime\prime}}|\xi_{t^\prime}|^2+e^{-\Gamma t^{\prime}}|\xi_{t^{\prime\prime}}|^2\right)\bigg\}\varrho_{ee}(0).
\end{eqnarray} 
Let us stress the above probability densities allow us completely characterize counts in the transmitted and reflected fields. Clearly, by taking the suitable integrals over chosen densities, we obtain the probabilities of particular events. For instance,
\begin{equation}
P_{R}(t)=\int_{0}^{t}dt^{\prime}p_{0}^{t}(R,t^{\prime}),
\end{equation}
is the probability of one count on the right side and no counts on the left side up to $t$. The probability that we observe two counts at the right detector and no counts at the left detector up to $t$ is given by
\begin{equation}
P_{RR}(t)=\int_{0}^{t}dt^{\prime\prime}\int_{0}^{t^{\prime\prime}}dt^{\prime}p_{0}^{t}\left(R,t^{\prime\prime};R,t^{\prime}\right)
\end{equation}  
Similarly we define $P_{L}(t)$, $P_{RL}(t), P_{LR}(t)$, and $P_{LL}(t)$. The probability that we do not detect any photon at the right detector up to $t$ is defined by
\begin{equation}
P_{0}^{t}(0)+P_{L}(t)+P_{LL}(t). 
\end{equation}

Let us note that having these probabilities, one can find the mean number of photons counted on the right side up to $t$, 
\begin{equation}
\langle N_{R}(t)\rangle = P_{R}(t)+P_{LR}(t)+P_{RL}(t)+2P_{RR}(t),
\end{equation}
and the mean number of photons counted on the left up to $t$,
\begin{equation}
\langle N_{L}(t)\rangle = P_{L}(t)+P_{LR}(t)+P_{RL}(t)+2P_{LL}(t).
\end{equation}

\subsection{Mean time of detections of photons in the output fields}

Making use of the exclusive probability densities, one can determine formulae for the probability densities of the times of successive counts.  Clearly, we take into account here the counts from the left and the right detector. The mean time of the first count can be calculated from the formula
\begin{equation}\label{tau1}
\tau_{1}=\int_{0}^{+\infty}dt\ t\ p_{1}(t)
\end{equation}
where $w_1(t)$, defined as
 \begin{equation}
 p_{1}(t)=-\frac{d}{dt}P_{0}^{t}(0),
 \end{equation}
 is the probability density that the first count is recorded around time $t$ (strictly in the interval $[t,t+dt)$) given that the detectors start measuring at time $t=0$. One can check that 
 \begin{eqnarray}
 p_{1}\left(t\right)&=& \left\{\left|\xi_{t}-\Gamma_{1}\int_{0}^{t}ds
 e^{\left(-i\Delta_{0}+\frac{\Gamma}{2}\right)\left(s-t\right)}\xi_{s}\right|^2
 \right.\nonumber\\
&& \left.+\Gamma_{1}\Gamma_{2}e^{-\Gamma t}
 \left|\int_{0}^{t}dse^{\left(-i\Delta_{0}+\frac{\Gamma}{2}\right)s}\xi_{s}\right|^2\right\}\rho_{gg}(0)\nonumber \\
 &&+e^{-\Gamma t}\left(\Gamma\int_{t}^{+\infty}ds|\xi_{s}|^2+|\xi_{t}|^2\right) \varrho_{ee}(0).
 \end{eqnarray}
Note that for the atom being initially in the ground state the quantity
\begin{equation}
\tau_{d}=\int_{0}^{+\infty}dt\ t\ |\xi_{t}|^2-\tau_{1}
\end{equation}
defines the mean time of the photon delay due to the interaction with the system. 

If the atom is initially in the excited state one can determine also the mean time of the second count given by  
\begin{equation}\label{tau2}
\tau_{2}=\int_{0}^{+\infty}dt\ t\ p_{2}(t)
\end{equation}
where
\begin{equation}
p_{2}(t)=\int_{0}^{t}dt^{\prime}p(t,t^{\prime})
\end{equation}
and
\begin{eqnarray}
p(t,t^{\prime})&=&
\Gamma_{1}
e^{-\Gamma(t^{\prime}+t)}\left| \xi_{t^{\prime}}e^{-\left(i\Delta_{0}-\frac{\Gamma}{2}\right)t^{\prime}}+\xi_{t}e^{-\left(i\Delta_{0}-\frac{\Gamma}{2}\right)t}
\right.\nonumber\\&&\left.-\Gamma 
\int_{t^{\prime}}^{t}ds  e^{\left(-i\Delta_{0}+\frac{\Gamma}{2}\right)s}\xi_{s}
\right|^2+\Gamma_{2}\left(e^{-\Gamma t}|\xi_{t^\prime}|^2+e^{-\Gamma t^{\prime}}|\xi_{t}|^2\right).
\end{eqnarray}  
One can check that
\begin{equation}
\int_{0}^{+\infty}dt\int_{0}^{t}dt^{\prime}p(t,t^{\prime})=1.
\end{equation}

\subsection{Exponential pulse}

We illustrate our results for a  decaying exponential pulse defined by
\begin{equation}
\xi_{t}=\sqrt{\Omega}\exp\left(-\frac{\Omega}{2}t\right),
\end{equation}
where $\Omega>0$. In this case, from (\ref{Pzero}), we obtain the probability of zero detections up to $t$ of the form 
\begin{eqnarray}\label{Pzeroexpo}
P_{t}(0)&=&e^{-(\Gamma+\Omega)t}\rho_{ee}(0)
+\left[e^{-\Omega t}+\frac{4\Omega\Gamma_1}{(\Gamma-\Omega)^2+4\Delta_{0}^2}
\right.\nonumber\\
&&\left.\times\left(e^{-\Omega t}+ e^{-\Gamma t}-2\cos \left(\Delta_{0}t\right) e^{-\frac{1}{2}(\Gamma+\Omega)t}\right)\right]\varrho_{gg}(0).
\end{eqnarray}
For the resonant case ($\Delta_{0}=0$) and $\Omega=\Gamma$ we have
\begin{equation}
P_{t}(0)=e^{-2\Gamma t} \varrho_{ee}(0)+e^{-\Gamma t}\left(1+\Gamma \Gamma_{1}t^2\right)\varrho_{gg}(0).
\end{equation} 
One can easily check that (\ref{Pzeroexpo}) reduces to
\begin{eqnarray}
P_{t}(0)&=&e^{-(\Gamma+\Omega)t}\rho_{ee}(0)
+e^{-\Omega t}\varrho_{gg}(0)
\end{eqnarray}
in the limit of $|\Delta_{0}|\to+\infty$. This expression reflects the fact that for large values of the detuning the input field is not able to drive the system and we deal with two independent sources of photons in the output field.

In the limit of the large time $t\to +\infty$, we get the following formulae 
\begin{equation}\label{l1}
\lim_{t\to+\infty}P_{R}(t)= \left[1-\frac{4\Gamma_{1}\Gamma_{2}\left(\Omega+\Gamma\right)}{\Gamma\left(4\Delta_{0}^2+\left(\Gamma+\Omega\right)^2\right)}\right]\varrho_{gg}(0),
\end{equation}
\begin{equation}\label{l2}
\lim_{t\to+\infty}P_{L}(t)= \frac{4\Gamma_{1}\Gamma_{2}\left(\Omega+\Gamma\right)\varrho_{gg}(0)}{\Gamma\left(4\Delta_{0}^2+\left(\Gamma+\Omega\right)^2\right)},
\end{equation}
\begin{eqnarray}\label{l3}
\lim_{t\to+\infty}P_{RR}(t)&=& \Gamma_{1}\left(4\Delta_{0}^2+\Gamma^2-4\Gamma\Gamma_1+6\Gamma\Omega+4\Gamma_1^2-4\Gamma_1\Omega+\Omega^2\right)\varrho_{ee}(0)
\nonumber\\
&&\times\Gamma^{-1}\left(4\Delta_{0}^2+\left(\Gamma+\Omega\right)^2\right)^{-1},
\end{eqnarray}
\begin{eqnarray}\label{l4}
\lim_{t\to+\infty}P_{LR}(t)&=& \left[\Gamma_{2}\left(4\Delta_{0}^2\Omega+4\Gamma_1^2\Omega-4\Gamma_1\Omega^2+\Omega^3\right)+\Gamma^2\Gamma_2\Omega \right.
\nonumber\\&&\left.+\Gamma\Gamma_2\left(4\Gamma_1^2-4\Gamma_1\Omega+2\Omega^2\right)\right]\varrho_{ee}(0)
\nonumber\\
&&\times\Gamma^{-1}\left(\Gamma+\Omega\right)^{-1}\left(4\Delta_{0}^2+\left(\Gamma+\Omega\right)^2\right)^{-1},
\end{eqnarray}
\begin{eqnarray}\label{l5}
\lim_{t\to+\infty}P_{RL}(t)&=&\Gamma_{2}\left(4\Delta_{0}^2\Gamma+\Gamma^3-4\Gamma^2\Gamma_1+2\Gamma^2\Omega+4\Gamma\Gamma_1^2\right.\nonumber\\&&\left.-4\Gamma\Gamma_1\Omega+\Gamma\Omega^2+4\Gamma_1^2\Omega\right)\varrho_{ee}(0)
\nonumber\\
&&\times\Gamma^{-1}\left(\Gamma+\Omega\right)^{-1}\left(4\Delta_{0}^2+\left(\Gamma+\Omega\right)^2\right)^{-1},
\end{eqnarray}
\begin{equation}\label{l6}
\lim_{t\to+\infty}P_{LL}(t)= \frac{4\Gamma_{1}\Gamma_{2}^2\varrho_{ee}(0)}{\Gamma\left(4\Delta_{0}^2+\left(\Gamma+\Omega\right)^2\right)}.
\end{equation}
If $|\Delta_{0}|\to \infty$ then (\ref{l1})-(\ref{l6}) have a simple interpretation. Namely, if the atom is initially in the ground state, we are certain that we observe one photon from the right. If the atom is prepared in the excited state, we may observe photons from the left and from the right, but we do not observe two photons from the left. Moreover, in this case for $\Gamma_{1}=\Gamma_{2}=0.5\Gamma$ and $\Gamma\ll \Omega$, we have $P_{RR}(+\infty)=0.5$, $P_{LR}(+\infty)=0.5$, and $P_{RL}(+\infty)=0$. Clearly, the assumption $\Gamma\ll \Omega$ means that the photon pulse is taken much shorter than the mean time of spontaneous decay of the atom. For $\Omega\ll \Gamma$, we get $P_{RR}(+\infty)=0.5$, $P_{LR}(+\infty)=0$, and $P_{RL}(+\infty)=0.5$. Note that in both situations the input photon is transmitted with the probability equal to one.  

In the resonant case for $\Gamma_{1}=\Gamma_{2}=0.5\Gamma$, we get from (\ref{l1})-(\ref{l6}) respectively
\begin{equation}\label{pr}
\lim_{t\to+\infty}P_{R}(t)=\frac{\Omega}{\Gamma+\Omega}\varrho_{gg}(0),
\end{equation} 
\begin{equation}\label{pl}
\lim_{t\to+\infty}P_{L}(t)=\frac{\Gamma}{\Gamma+\Omega}\varrho_{gg}(0), \end{equation} 
\begin{equation}\label{prr}
\lim_{t\to+\infty}P_{RR}(t)= \frac{\Omega\left(\Omega+4\Gamma\right)}{2\left(\Gamma+\Omega\right)^2}\varrho_{ee}(0),
\end{equation}
\begin{equation}\label{plr}
\lim_{t\to+\infty}P_{LR}(t)=\left(\frac{1}{2}- \frac{3\Gamma\Omega}{2\left(\Gamma+\Omega\right)^2}\right)\varrho_{ee}(0),
\end{equation}
\begin{equation}\label{prl}
\lim_{t\to+\infty}P_{RL}(t)=\frac{\Gamma\Omega}{2\left(\Gamma+\Omega\right)^2}\varrho_{ee}(0),
\end{equation}
\begin{equation}\label{pll}
\lim_{t\to+\infty}P_{LL}(t)= \frac{\Gamma^2}{2\left(\Gamma+\Omega\right)^2}\varrho_{ee}(0),
\end{equation}
which agree with the results provided in the real-space approach for an infinite waveguide both for the atom initially in the ground state \cite{Fan2010,Nysteen2015} and the excited state \cite{Fan2012}. 
From (\ref{pr})-(\ref{pll}), we obtain then
\begin{equation}
\lim_{t\to+\infty}\langle N_{R}(t)\rangle =\frac{\Omega}{\Gamma+\Omega}\rho_{gg}(0)+ \frac{\Gamma^2 + 8\Gamma\Omega + 3\Omega^2}{2\left(\Gamma + \Omega\right)^2}\varrho_{ee}(0),
\end{equation}
\begin{equation}
\lim_{t\to+\infty}\langle N_{L}(t)\rangle =\frac{\Gamma}{\Gamma+\Omega} \rho_{gg}(0)+ \frac{3\Gamma^2 + \Omega^2}{2\left(\Gamma + \Omega\right)^2}\varrho_{ee}(0).
\end{equation}
Then if $\Gamma\ll\Omega$, we obtain $P_{R}(+\infty)=\varrho_{gg}(0)$, $P_{L}(+\infty)=0$, $P_{RR}(+\infty)=0.5\varrho_{ee}(0)$, $P_{LR}(+\infty)=0.5\varrho_{ee}(0)$, $P_{RL}(+\infty)=P_{LL}(+\infty)=0$, $\langle N_{R}(+\infty)\rangle=\varrho_{gg}(0)+1.5\varrho_{ee}(0)$, and $\langle N_{L}(+\infty)\rangle=0.5\varrho_{ee}(0)$. 
Thus, it is seen that for the photon pulse much shorter than the spontaneous emission lifetime of the atom, the input photon is completely transmitted. Note that if $\Omega\ll\Gamma$, we obtain $P_{R}(+\infty)=0$, $P_{L}(+\infty)=\varrho_{gg}(0)$, $P_{RR}(+\infty)=0$, $P_{LR}(+\infty)=0.5\varrho_{ee}(0)$, $P_{RL}(+\infty)=0$, $P_{LL}(+\infty)=0.5\varrho_{ee}(0)$, $\langle N_{R}(+\infty)\rangle=0.5\varrho_{ee}(0)$, and $\langle N_{L}(+\infty)\rangle=\rho_{gg}(0)+1.5\varrho_{ee}(0)$. Then the input photon is perfectly reflected. 

Now we analyse the mean time of the detections. From (\ref{tau1}), we obtain the formula
\begin{equation}
\tau_{1}=\frac{1}{\Gamma+\Omega}\varrho_{ee}(0)
+\left(\frac{1}{\Omega}+\frac{4\Gamma_{1}\left(\Omega+\Gamma\right)}{\Gamma\left(4\Delta_{0}^2+\left(\Gamma+\Omega\right)^2\right)}\right)\varrho_{gg}(0),
\end{equation}
which in the resonance for $\Gamma_1=\Gamma_2=0.5\Gamma$ reduces to
\begin{equation}
\tau_1=\frac{1}{\Gamma+\Omega} \rho_{ee}(0)+\frac{\Gamma+3\Omega}{\Omega\left(\Gamma+\Omega\right)}\varrho_{gg}(0).
\end{equation}

If initially the atom is in the excited state, we can determine the mean time of detection of the second photon. From (\ref{tau2}) for the exponential pulse, we find that
\begin{eqnarray}
\tau_2&=&\left[4\Delta_{0}^2\left(\Gamma^2+\Gamma\Omega+\Omega^2\right)+\Gamma^4+3\Gamma^3\Omega+4\Gamma^2\Omega^2+4\Gamma_1\Gamma^2\Omega
\right.\nonumber\\&&\left.+3\Gamma\Omega^3-4\Gamma_1\Omega^3+\Omega^4\right]
\Gamma^{-1}\Omega^{-1}(\Gamma+\Omega)^{-1}\left(4\Delta_{0}^2+(\Gamma+\Omega)^2\right)^{-1}.
\end{eqnarray}

\begin{figure}[h]
 	\includegraphics[width=8cm,height=6cm]{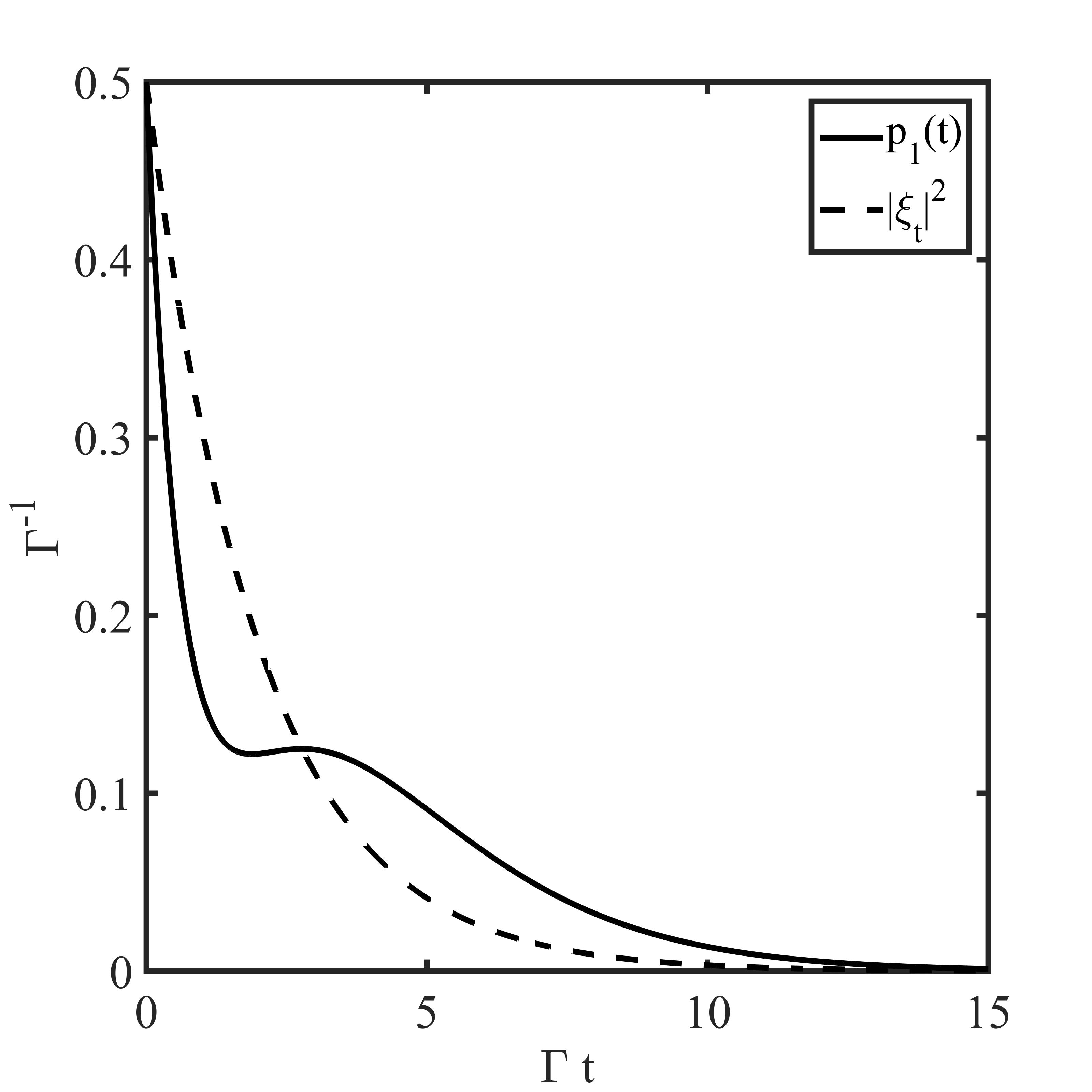}
 	\includegraphics[width=8cm,height=6cm]{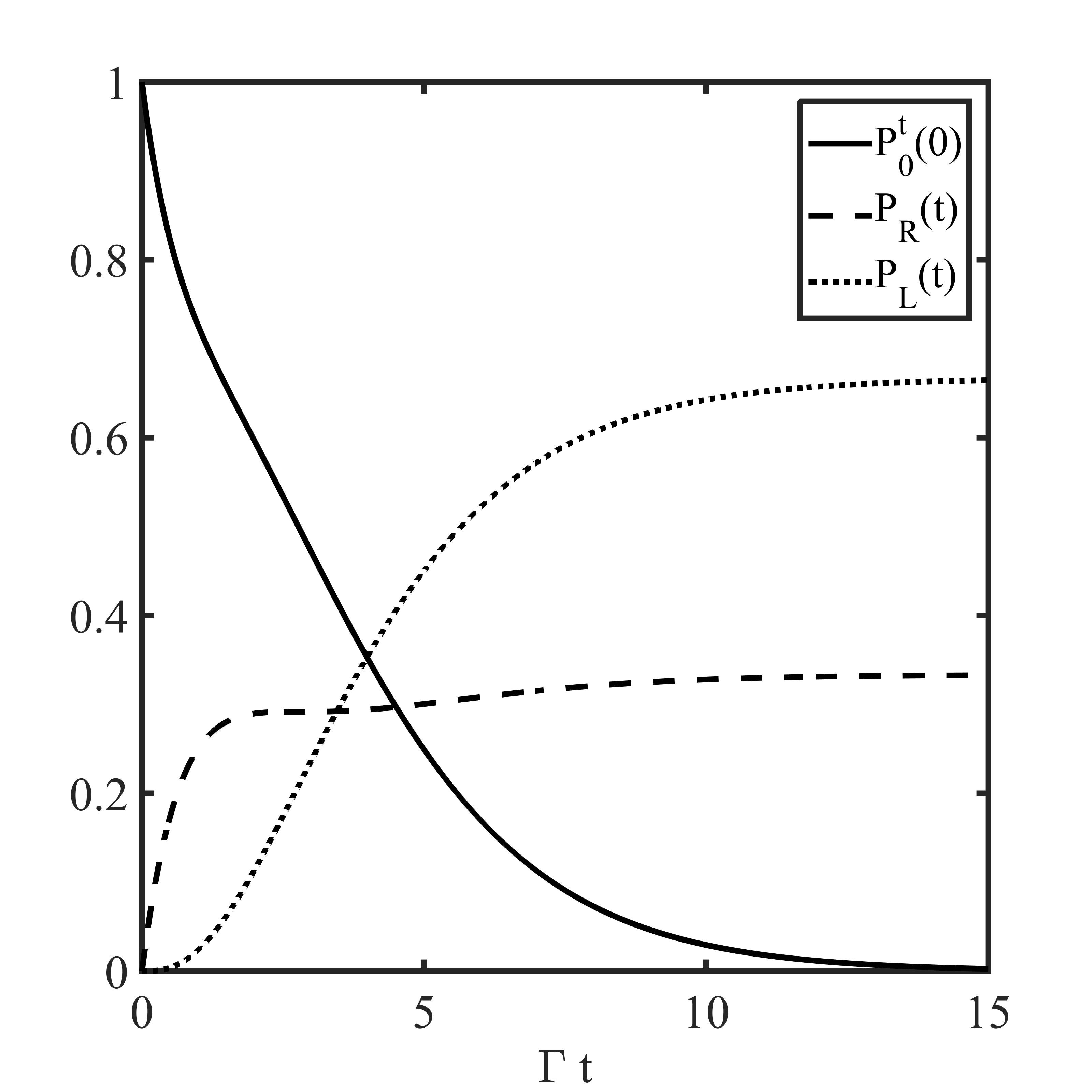}
 	\caption{Photon counting characteristics of the output field for $\rho_{gg}(0)=1$, $\Gamma=1$, $\Delta_{0}=0$, and $\Omega=0.5$. Here $\tau_1=3.33\Gamma^{-1}$.} \label{Fig-ground}
 \end{figure}

\begin{figure}[h]
 	\includegraphics[width=8cm,height=6cm]{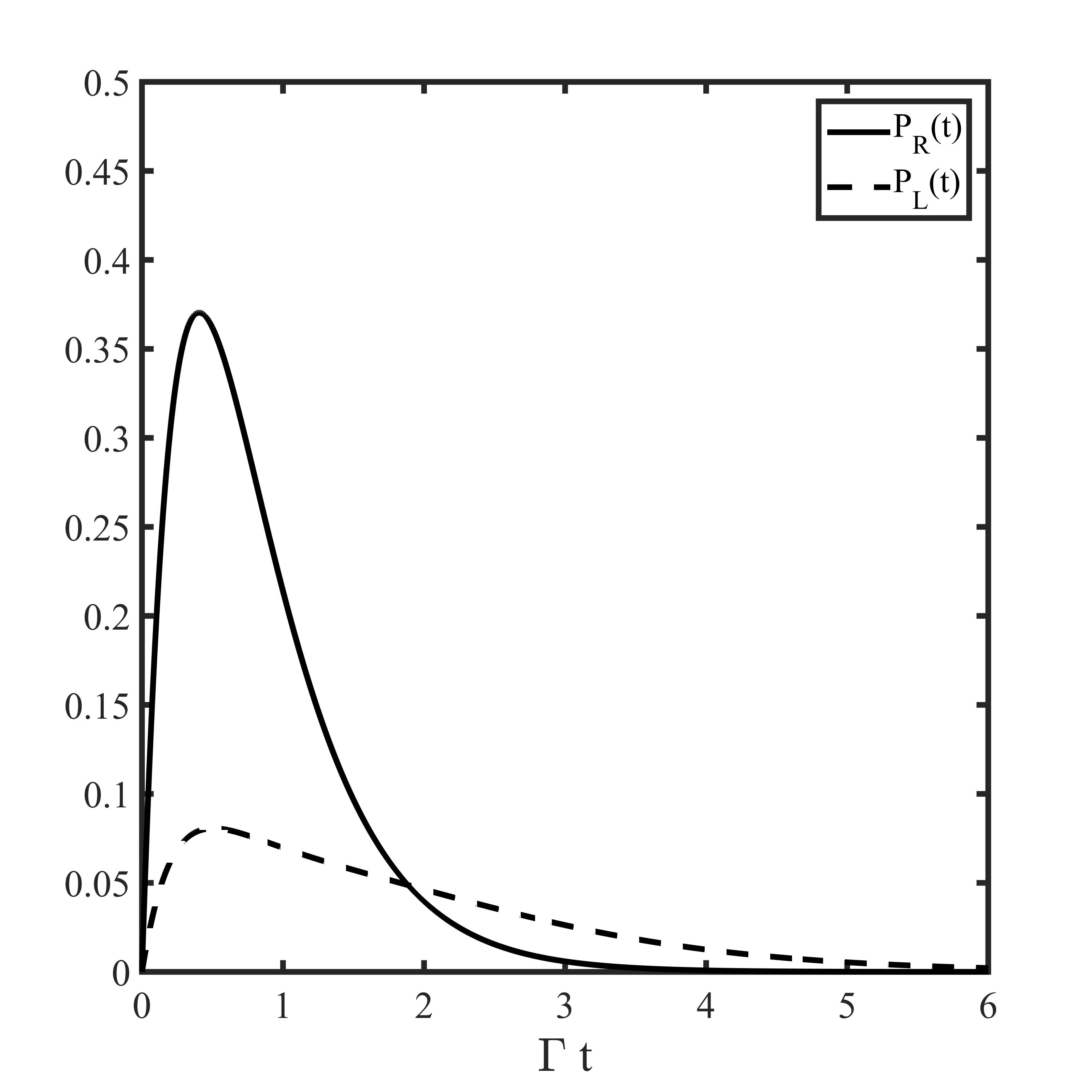}
 	\includegraphics[width=8cm,height=6cm]{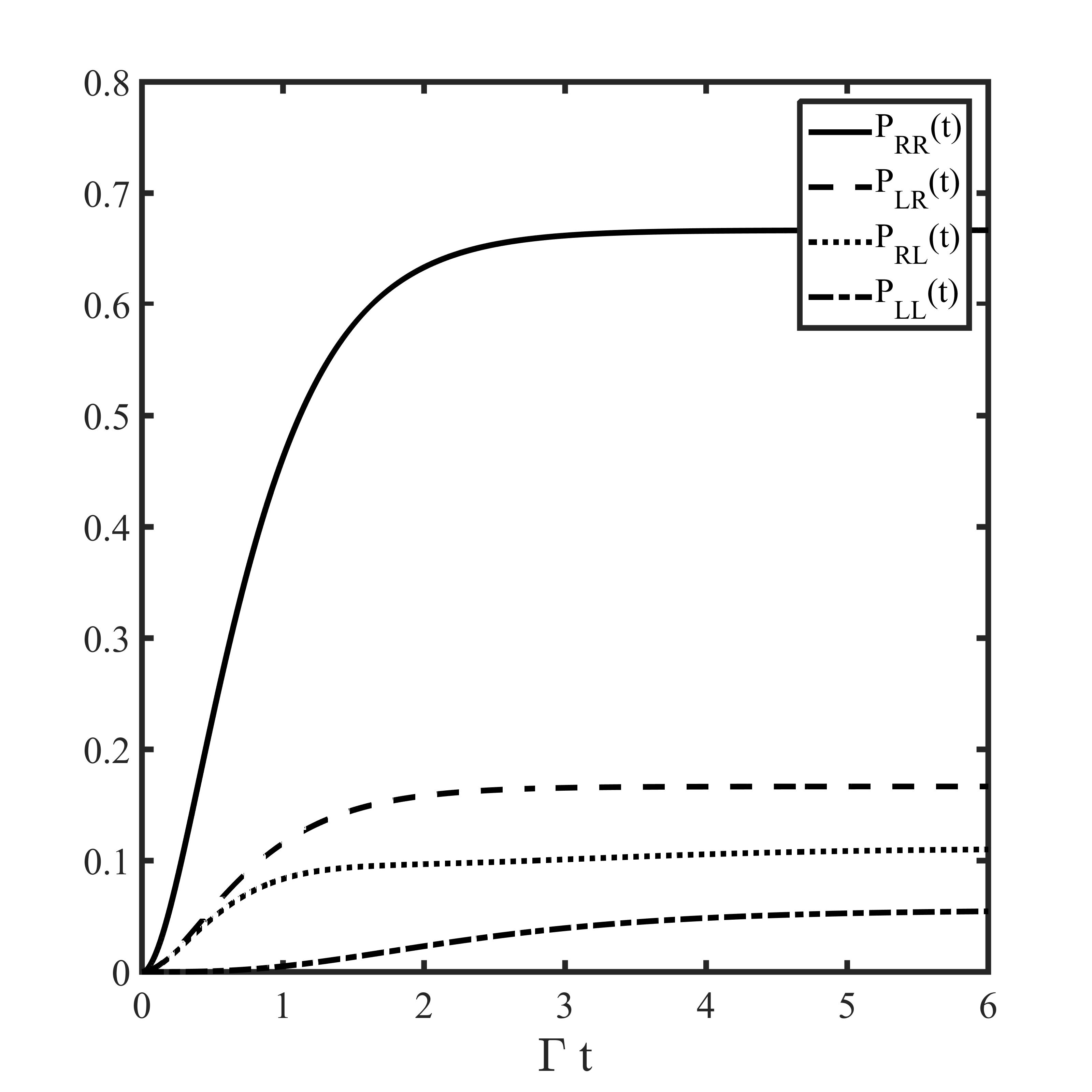}
 	\includegraphics[width=8cm,height=6cm]{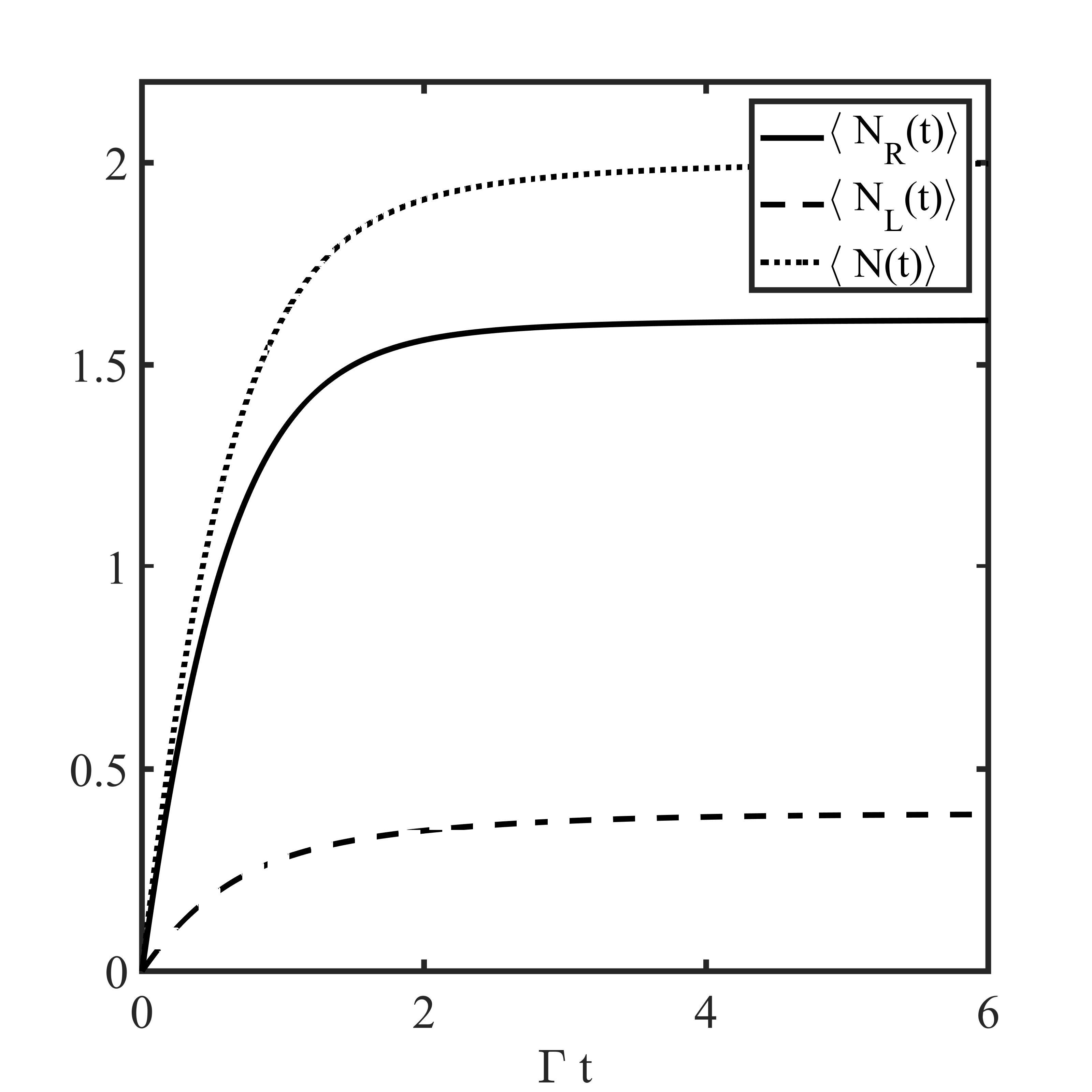}
 		\includegraphics[width=8cm,height=6cm]{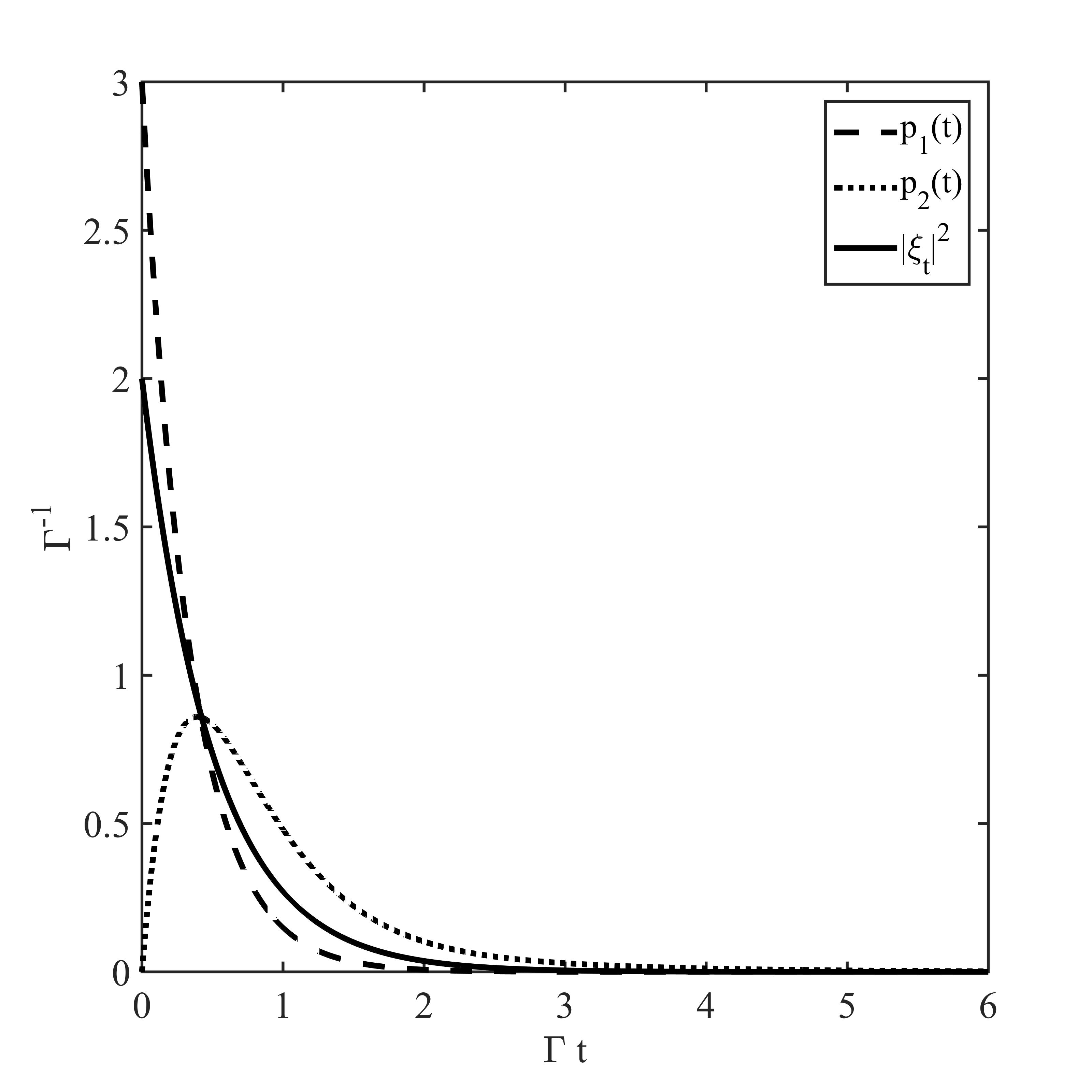}
 	\caption{Photon counting characteristics of the output field for $\rho_{ee}(0)=1$, $\Gamma=1$, $\Delta_{0}=0$, and $\Omega=2$. Here $\tau_{1}=0.33\Gamma^{-1}$, $\tau_2=0.94\Gamma^{-1}$.} \label{Fig-excited}
\end{figure}

For $\Delta_{0}=0$ and $\Gamma_{1}=\Gamma_2=0.5\Gamma$ we get
\begin{equation}
\tau_2=\frac{\Gamma^3+4\Gamma^2\Omega+\Omega^3}{\Gamma\Omega\left(\Gamma+\Omega\right)^2}.
\end{equation} 
Let us note that for $\Gamma\ll \Omega$ we obtain the expected values $\tau_1=1/\Omega$, $\tau_2=1/\Gamma$, and for $\Omega\ll\Gamma$ we get $\tau_1=1/\Gamma$ and $\tau_2=1/\Omega$. The properties of the output field for the intermediate values of the parameters are presented for the resonant case in Figs. \ref{Fig-ground} and \ref{Fig-excited}. The characteristics of the output field for the atom being initially in the ground state are depicted in Fig. \ref{Fig-ground}. Note that in this case $P_{R}(t)$ and $P_{L}(t)$ are equal to the mean number of photons counted, respectively, from the right and the left side up to $t$. The ratio of the transmitted and reflected light is expressed then via $\Omega/\Gamma$. The mean time of the photon delay is $\tau_{d}=2/(\Gamma+\Omega)$.  Fig. \ref{Fig-excited} shows the features of the output field if the atom is fully excited initially.

\section{Conclusion}

Using quantum filtering theory we have solved the problem of scattering of a single-photon field on a quantum system. In order to determine the photon counting probabilities for the transmitted and reflected fields, we have decomposed the reduced state of the system interacting with light into the sum of integrals over quantum trajectories associated with the two-dimensional counting stochastic process defined for the bidirectional output field.
We have derived analytical formulae for the quantum trajectories by making use of the collision model wherein the evolution of the quantum system is defined by discrete in time sequence of collisions of the system with bath ancillas (qubits). We have determined the stochastic representation of the reduced state of the quantum system by considering the repeated measurements performed on qubits constituting the output field. We have shown that in the continuous-time limit the discrete approach yields filtering and master equations sets, which are consistent with the equations obtained via quantum stochastic calculus \cite{Gough12a,Gough12b,Baragiola2012,Gough13,Baragiola17}. We have determined analytical formulae for quantum trajectories expressing them by  conditional vectors which offer a simple and intuitive physical interpretation. Finally, we have applied our results to describe the scattering of a single-photon field on a two-level atom.  We have found analytical formulae for photon counting probabilities for an arbitrary photon profile and any initial state of the atom. Of course, our results allow one to derive also the asymptotic probabilities of transition and reflections, and we have checked that they agree with the formulae obtained in  real-space approach for an exponential pulse \cite{Fan2010,Fan2012}. We have derived, moreover, general formulae for the probability densities of times of the successive counts. We have also found analytic expressions for the mean time of the first and second counts for an exponential pulse. We would like to emphasize that the presented method can be effectively applied for a three-level atom or cavity mode, and it can be extended for studying the scattering of light in some other states, in particular in $N$-photon state. We have formulated the problem in one dimension, however, its generalization to higher dimensions is straightforward although nontrivial. We would like to stress that even this simple model can be tested experimentally, see, for instance, \cite{Scarani2013,Leong2016}.

\section*{Acknowledgements}  
This research was supported by the National Science Centre project 2018/30/A/ST2/00837.

\appendix
\section{The interaction}\label{V}
Writing down the unitary operator, ${V}_{k}$, defined by (\ref{operatorV}), in the representation of $\{|00\rangle_{k}, |01\rangle_{k}, |10\rangle_{k}, |11\rangle_{k}\}$, we get the following system's operators 
\begin{eqnarray}\label{vmatrix1}
V_{00,00}&=& \mathbbm{1}\!-\!i\tau \left[H_{\mathcal{S}}\!-\!  \frac{i}{2}\left(L^\dagger_{1} L_{1}\!+\!L^\dagger_{2} L_{2}\right) \right]\!+\!O(\tau^{2})\!,\\
V_{00,01}&=&V_{10,11}=-\sqrt{\tau} L^{\dagger}_{2} + O(\tau^{3/2}), \\
V_{00,10}&=&V_{01,11}=- \sqrt{\tau} L^\dagger_{1} + O(\tau^{3/2}),\\
V_{01,00}&=& V_{11,10}=\sqrt{\tau}L_{2}\!+\!O(\tau^{3/2}),\\
V_{01,01}&=& \mathbbm{1}\!-\! i\tau \left[H_{\mathcal{S}}\!-\!  \frac{i}{2}\left(L^\dagger_{1} L_{1}\!+\!L_{2} L_{2}^{\dagger}\right) \right]\!+\!O(\tau^{2})\!,\\
V_{10,00}&=&V_{11,01}= \sqrt{\tau} L_{1} + O(\tau^{3/2}),\\
V_{10,10}&=&\mathbbm{1}\!-\!i\tau \left[H_{\mathcal{S}}\!-\!  \frac{i}{2}\left(L_{1}L_{1}^{\dagger}\!+\!L_{2}^{\dagger}L_{2}\right) \right]\!+\!O(\tau^{2})\!,\\
V_{11,11}&=& \mathbbm{1}\!-\!i\tau \left[H_{\mathcal{S}}\!-\!  \frac{i}{2}\left(L_{1} L_{1}^{\dagger}\!+\!L_{2}L_{2}^{\dagger} \right) \right]\!+\!O(\tau^{2})\!,\\
V_{11,00}&=&\frac{\tau}{2}\left(L_{1}L_{2}+L_{2}L_{1}\right)+O(\tau^2),\\
V_{00,11}&=&\frac{\tau}{2}\left(L_{1}^{\dagger}L_{2}^{\dagger}+L_{2}^{\dagger}L_{1}^{\dagger}\right)+O(\tau^2),\\
V_{01,10}&=&-\frac{\tau}{2}\left(L_{1}^{\dagger}L_{2}+L_{2}L_{1}^{\dagger}\right)+O(\tau^2),\\
V_{10,01}&=&-\frac{\tau}{2}\left(L_{1}L_{2}^{\dagger}+L_{2}^{\dagger}L_{1}\right)+O(\tau^2).\label{vmatrix2}
\end{eqnarray}
\section{Proof to the Theorem \ref{TH-1}}\label{proof}

We proof the Theorem (\ref{TH-1}) by induction. We show that if (\ref{cond2}) holds for any given case $j$, then it also holds for $j+1$. First we observe that $|\Psi_{j|  \pmb{\eta}_j} \rangle$ can be written in the form
\begin{eqnarray}
|\Psi_{j|  \pmb{\eta}_j} \rangle &=& |0\rangle_{j} \otimes  \sum_{k=j+1}^{N-1} \sqrt{\tau}\xi_k|1_{k}\rangle_{[j+1} \otimes |vac \rangle_{[j} \otimes|\alpha_{j|  \pmb{\eta}_j}\rangle  \nonumber \\ &&+  | vac \rangle_{[j} \otimes|vac \rangle_{[j} \otimes |\beta_{j|  \pmb{\eta}_j}\rangle \nonumber \\
&&+ |1\rangle_{j}\otimes |vac\rangle_{[j+1}\otimes |vac \rangle_{[j} \otimes \sqrt{\tau}\xi_{j}|\alpha_{j|  \pmb{\eta}_j}\rangle.
\end{eqnarray}
Now acting by the unitary operator ${V}_{j}$ on $|\Psi_{j|\pmb{\eta}_j} \rangle$ one finds
	\begin{eqnarray}\label{cond3}
	 {V}_{j} |\Psi_{j| \pmb{\eta}_j} \rangle &=& |0\rangle_{j} \otimes\sum_{k=j+1}^{N-1} \sqrt{\tau}\xi_{k} |1_{k}\rangle_{[j+1}\otimes|vac \rangle_{[j}   \otimes V_{00,00}|\alpha_{j|  \pmb{\eta}_j} \rangle \nonumber\\
	&&+ | vac \rangle_{[j} \otimes|vac \rangle_{[j} \otimes 
	\left( V_{00,00}|\beta_{j|  \pmb{\eta}_j} \rangle + \sqrt{\tau}\xi_{j}V_{00,10}|\alpha_{j|  \pmb{\eta}_j} \rangle \right) \nonumber\\ && 
	+ |1\rangle_{j} \otimes \sum_{k=j+1}^{N-1} \sqrt{\tau}\xi_{k} |1_{k}\rangle_{[j+1} \otimes | vac \rangle_{[j} \otimes V_{10,00}|\alpha_{j|  \pmb{\eta}_j} \rangle \nonumber 
	\\&&+|1\rangle_{j}\otimes | vac \rangle_{[j+1} \otimes | vac \rangle_{[j} \otimes\left(V_{10,00}|\beta_{j|  \pmb{\eta}_j} \rangle+ \sqrt{\tau}\xi_{j}V_{10,10}|\alpha_{j|  \pmb{\eta}_j} \rangle\right)\nonumber\\
	&&+ |0\rangle_{j} \otimes\sum_{k=j+1}^{N-1} \sqrt{\tau}\xi_{k} |1_{k}\rangle_{[j+1} \otimes|1\rangle_{j} \otimes | vac \rangle_{[j+1} \otimes V_{01,00}|\alpha_{j|  \pmb{\eta}_j} \rangle \nonumber 
	\\&&+ | vac \rangle_{[j} \otimes|1\rangle_{j} \otimes | vac \rangle_{[j+1} \otimes \left(V_{01,00}|\beta_{j|  \pmb{\eta}_j} \rangle+\sqrt{\tau}\xi_{j}V_{01,10}|\alpha_{j|  \pmb{\eta}_j} \rangle\right) \nonumber
	\\&& + |1\rangle_{j}\otimes \sum_{k=j+1}^{N-1} \sqrt{\tau}\xi_{k} |1_{k}\rangle_{[j+1}\otimes |1\rangle_{j}\otimes| vac \rangle_{[j+1} \otimes V_{11,00}|\alpha_{j|  \pmb{\eta}_j} \rangle \nonumber 
	\\&&+|1\rangle_{j}\otimes| vac \rangle_{[j+1} \otimes |1\rangle_{j}\otimes| vac \rangle_{[j+1} \otimes
	\left(V_{11,00}|\beta_{j|  \pmb{\eta}_j} \rangle+\sqrt{\tau}\xi_{j}V_{11,10}|\alpha_{j|  \pmb{\eta}_j} \rangle\right).
	\end{eqnarray}
For the measurement of the two observables 
\begin{equation}
\sigma_{l,j}^{+}\sigma_{l,j}^{-},\;\;l=1,2
\end{equation}
we define the conditional vector $|\Psi_{j+1|  \pmb{\eta}_{j+1}} \rangle$ in   $\mathcal{H}_{\mathcal{E}}^{[j+1}\otimes \mathcal{H}_{S}$ by 
\begin{equation}
\Pi_{\eta_{j+1}} {V}_{j}|\Psi_{j|  \pmb{\eta}_j} \rangle=|\eta_{j+1}\rangle_{j}\otimes|\Psi_{j+1|  \pmb{\eta}_{j+1} }\rangle,
\end{equation}
where $\eta_{j+1}=00,01, 10, 11$  denotes four results of the measurement and
\begin{eqnarray}
\Pi_{00}=|00\rangle_{j}{}_{j}\langle 00|,\;\;\;\Pi_{01}=|01\rangle_{j}{}_{j}\langle 01|,
\end{eqnarray}
\begin{eqnarray}
\Pi_{10}=|10\rangle_{j}{}_{j}\langle 10|,\;\;\;\Pi_{11}=|11\rangle_{j}{}_{j}\langle 11|,
\end{eqnarray}
Hence finally one gets
\begin{eqnarray}
|\Psi_{j+1|  \pmb{\eta}_{j+1} }\rangle &=& \sum_{k=j+1}^{N-1} \sqrt{\tau}\xi_{k} |1_k \rangle_{[j+1} \otimes | vac \rangle_{[j+1} \otimes |\alpha_{j+1| \pmb{\eta}_{j+1} }\rangle \nonumber\\ &&+ | vac \rangle_{[j+1} \otimes | vac \rangle_{[j+1} \otimes |\beta_{j+1|  \pmb{\eta}_{j+1} }\rangle
\end{eqnarray}
with  $|\alpha_{j+1|  \pmb{\eta}_j+1}\rangle$, $|\beta_{j+1|  \pmb{\eta}_j+1}\rangle$  satisfying recurrent Eqs. (\ref{rec1a})-(\ref{rec4b}), which ends the proof.

\section{Discrete conditional vectors}\label{condivectors}

We present below solutions to Eqs. (\ref{rec1a})-(\ref{rec3b}) for some chosen sequences of the outcomes:

\begin{enumerate}
	
	\item $\pmb{\eta}_j=\mathbf{0}_j=\left((0,0)_{j},\ldots, (0,0)_{1}\right)$, so there is no any count from $0$ up to $j\tau$:
	
	\begin{equation}\label{convec1a}
	|\alpha_{j|\mathbf{0}_j}\rangle=V_{00,00}^{j}|\psi_{0}\rangle,
	\end{equation}
	\begin{equation}\label{convec1b}
	|\beta_{j|\mathbf{0}_j}\rangle=  \sum_{k=0}^{j-1}V_{00,00}^{j-k-1}\sqrt{\tau}\xi_{k}V_{00,10}V_{00,00}^k|\psi_{0}\rangle.
	\end{equation}
	
	\item one count at $\tau l_1$ at the right detector and no other counts from $0$ up to $j\tau$:
	
	\begin{equation}\label{convec2a}
	|\alpha_{j|R,l_1}\rangle=V_{00,00}^{j-l_{1}}V_{10,00}V_{00,00}^{l_{1}-1}|\psi_{0}\rangle,
	\end{equation}
	and
	\begin{eqnarray}\label{convec2b}
	|\beta_{j|R,l_1}\rangle&=&\bigg[V_{00,00}^{j-l_{1}}V_{10,00}\sum_{k=0}^{l_{1}-2}V_{00,00}^{l_{1}-k-2}\sqrt{\tau}\xi_{k}V_{00,10}V_{00,00}^k \nonumber 
	\\ &&+
	V_{00,00}^{j-l_{1}}	
	\sqrt{\tau}\xi_{l_{1}-1}V_{10,10}V_{00,00}^{l_{1}-1}\nonumber\\ 
	&&+
	\sum_{k=l_{1}}^{j-1}V_{00,00}^{j-k-1}	
	\sqrt{\tau}\xi_{k}V_{00,10}V_{00,00}^{k-l_{1}}V_{10,00}V_{00,00}^{l_{1}-1}	\bigg]|\psi_{0}\rangle. 
	\end{eqnarray}
	
	\item two counts at $\tau l_1$ and $\tau l_2$  both at the right detector and no other counts from $0$ up to $j\tau$:
	
	\begin{equation}\label{convec3a}
	|\alpha_{j|R,l_{2};R,l_{1}}\rangle=V_{00,00}^{j-l_{2}}V_{10,00}V_{00,00}^{l_{2}-l_{1}-1}V_{10,00}V_{00,00}^{l_{1}-1}|\psi_{0}\rangle,
	\end{equation}
	and
	\begin{eqnarray}\label{convec3b}
	|\beta_{j|R,l_{2};R,l_{1}}\rangle&=& \sqrt{\tau}
	\left[V_{00,00}^{j-l_{2}}V_{10,00}V_{00,00}^{l_{2}-l_{1}-1}\xi_{l_{1}-1}V_{10,10}V_{00,00}^{l_{1}-1}\right.\nonumber\\
	&&+V_{00,00}^{j-l_{2}}\xi_{l_{2}-1}V_{10,10}V_{00,00}^{l_{2}-l_{1}-1}V_{10,00}V_{00,00}^{l_{1}-1}\nonumber \\
	&&+V_{00,00}^{j-l_{2}}V_{10,00}V_{00,00}^{l_{2}-l_{1}-1}V_{10,00}\sum_{k=0}^{l_{1}-2}V_{00,00}^{l_{1}-k-2}\xi_{k}V_{00,10}V_{00,00}^k\nonumber\\
	&&+V_{00,00}^{j-l_{2}}V_{10,00}\sum_{k=l_{1}}^{l_{2}-2}V_{00,00}^{l_{2}-k-2}\xi_{k}V_{00,10}V_{00,00}^{k-l_{1}}V_{10,00}V_{00,00}^{l_{1}-1}\\
	&&\left.+\sum_{k=l_{2}}^{j-1}V_{00,00}^{j-k-1}\xi_{k}V_{00,10}V_{00,00}^{k-l_{2}}V_{10,00}V_{00,00}^{l_{2}-l_{1}-1}V_{10,00}V_{00,00}^{l_{1}-1}
	\right]|\psi_{0}\rangle \nonumber
	\end{eqnarray}
	
	\item one count at $\tau l_1$ at the left detector and no other counts from $0$ up to $j\tau$:
	
	\begin{equation}\label{convec4a}
	|\alpha_{j|L,l_1}\rangle=V_{00,00}^{j-l_{1}}V_{01,00}V_{00,00}^{l_{1}-1}|\psi_{0}\rangle,
	\end{equation}
	and
	\begin{eqnarray}\label{convec4b}
	|\beta_{j|L,l_1}\rangle&=& 
	\bigg[V_{00,00}^{j-l_{1}}V_{01,00}\sum_{k=0}^{l_{1}-2}V_{00,00}^{l_{1}-k-2}\sqrt{\tau}\xi_{k}V_{00,10}V_{00,00}^k 
	\nonumber 
	\\&&+V_{00,00}^{j-l_{1}}\sqrt{\tau}\xi_{l_{1}-1}V_{01,10}V_{00,00}^{l_{1}-1}	\nonumber 
	\\&&+ \sum_{k=l_{1}}^{j-1}V_{00,00}^{j-k-1}	
	\sqrt{\tau}\xi_{k}V_{00,10}V_{00,00}^{k-l_{1}}V_{01,00}V_{00,00}^{l_{1}-1}	\bigg]|\psi_{0}\rangle.
	\end{eqnarray}

	\item two counts at $\tau l_1$ and $\tau l_2$  both at the left detector and no other counts from $0$ up to $j\tau$:
	
	\begin{equation}\label{convec5a}
	|\alpha_{j|L,l_{2};L,l_{1}}\rangle=V_{00,00}^{j-l_{2}}V_{01,00}V_{00,00}^{l_{2}-l_{1}-1}V_{01,00}V_{00,00}^{l_{1}-1}|\psi_{0}\rangle,
	\end{equation}
	and
	\begin{eqnarray}\label{convec5b}
	|\beta_{j|L,l_{2};L,l_{1}}\rangle &=& 
	\left[V_{00,00}^{j-l_{2}}V_{01,00}V_{00,00}^{l_{2}-l_{1}-1}\sqrt{\tau}\xi_{l_{1}-1}V_{01,10}V_{00,00}^{l_{1}-1}\right.\nonumber\\
	&&+V_{00,00}^{j-l_{2}}\sqrt{\tau}\xi_{l_{2}-1}V_{01,10}V_{00,00}^{l_{2}-l_{1}-1}V_{01,00}V_{00,00}^{l_{1}-1}\\
	&&+V_{00,00}^{j-l_{2}}V_{01,00}V_{00,00}^{l_{2}-l_{1}-1}V_{01,00}\sum_{k=0}^{l_{1}-2}V_{00,00}^{l_{1}-k-2}\xi_{k}V_{00,10}V_{00,00}^k\nonumber\\
	&&+V_{00,00}^{j-l_{2}}V_{01,00}\sum_{k=l_{1}}^{l_{2}-2}V_{00,00}^{l_{2}-k-2}\xi_{k}V_{00,10}V_{00,00}^{k-l_{1}}V_{01,00}V_{00,00}^{l_{1}-1}
	\nonumber\\
	&&\left.+\sum_{k=l_{2}}^{j-1}V_{00,00}^{j-k-1}\xi_{k}V_{00,10}V_{00,00}^{k-l_{2}}V_{01,00}V_{00,00}^{l_{2}-l_{1}-1}V_{01,00}V_{00,00}^{l_{1}-1}
	\right]|\psi_{0}\rangle \nonumber
	\end{eqnarray}
	\item two counts at $\tau l_1$ and $\tau l_2$  the first one at the right detector and the second one in the left detector, and no other counts from $0$ up to $j\tau$:
	
	\begin{equation}\label{convec6a}
	|\alpha_{j|L,l_{2};R,l_{1}}\rangle=V_{00,00}^{j-l_{2}}V_{01,00}V_{00,00}^{l_{2}-l_{1}-1}V_{10,00}V_{00,00}^{l_{1}-1}|\psi_{0}\rangle,
	\end{equation}
	and
	\begin{eqnarray}\label{convec6b}
	|\beta_{j|L,l_{2};R,l_{1}}\rangle &=& \sqrt{\tau}
	\left[V_{00,00}^{j-l_{2}}V_{01,00}V_{00,00}^{l_{2}-l_{1}-1}\xi_{l_{1}-1}V_{10,10}V_{00,00}^{l_{1}-1}\right.\nonumber\\
	&&+V_{00,00}^{j-l_{2}}\xi_{l_{2}-1}V_{01,10}V_{00,00}^{l_{2}-l_{1}-1}
	V_{10,00}V_{00,00}^{l_{1}-1}\\
	&&
	+V_{00,00}^{j-l_{2}}V_{01,00}V_{00,00}^{l_{2}-l_{1}-1}V_{10,00}\sum_{k=0}^{l_{1}-2}V_{00,00}^{l_{1}-k-2}\xi_{k}V_{00,10}V_{00,00}^k\nonumber\\
	&&+V_{00,00}^{j-l_{2}}V_{01,00}\sum_{k=l_{1}}^{l_{2}-2}V_{00,00}^{l_{2}-k-2}\xi_{k}V_{00,10}V_{00,00}^{k-l_{1}}V_{10,00}V_{00,00}^{l_{1}-1}
	\nonumber\\
	&&\left.+\sum_{k=l_{2}}^{j-1}V_{00,00}^{j-k-1}\xi_{k}V_{00,10}V_{00,00}^{k-l_{2}}V_{01,00}V_{00,00}^{l_{2}-l_{1}-1}V_{10,00}V_{00,00}^{l_{1}-1}
	\right]|\psi_{0}\rangle \nonumber
	\end{eqnarray}
	
	\item two counts at $\tau l_1$ and $\tau l_2$  the first one at the left detector and the second one at the right detector, and no other counts from $0$ up to $j\tau$:
	
	\begin{equation}\label{convec7a}
	|\alpha_{j|R,l_{2};L,l_{1}}\rangle=V_{00,00}^{j-l_{2}}V_{10,00}V_{00,00}^{l_{2}-l_{1}-1}V_{01,00}V_{00,00}^{l_{1}-1}|\psi_{0}\rangle,
	\end{equation}
	and
	\begin{eqnarray}\label{convec7b}
	|\beta_{j|R,l_{2};L,l_{1}}\rangle &=& \sqrt{\tau}
	\left[V_{00,00}^{j-l_{2}}V_{10,00}V_{00,00}^{l_{2}-l_{1}-1}\xi_{l_{1}-1}V_{01,10}V_{00,00}^{l_{1}-1}\right.\nonumber\\
	&&+V_{00,00}^{j-l_{2}}\xi_{l_{2}-1}V_{10,10}V_{00,00}^{l_{2}-l_{1}-1}V_{01,00}V_{00,00}^{l_{1}-1}\\
	&&+V_{00,00}^{j-l_{2}}V_{10,00}V_{00,00}^{l_{2}-l_{1}-1}V_{01,00}\sum_{k=0}^{l_{1}-2}V_{00,00}^{l_{1}-k-2}\xi_{k}V_{00,10}V_{00,00}^k\nonumber\\
	&&+V_{00,00}^{j-l_{2}}V_{10,00}\sum_{k=l_{1}}^{l_{2}-2}V_{00,00}^{l_{2}-k-2}\xi_{k}V_{00,10}V_{00,00}^{k-l_{1}}V_{01,00}V_{00,00}^{l_{1}-1}
	\nonumber\\
	&&\left.+\sum_{k=l_{2}}^{j-1}V_{00,00}^{j-k-1}\xi_{k}V_{00,10}V_{00,00}^{k-l_{2}}V_{10,00}V_{00,00}^{l_{2}-l_{1}-1}V_{01,00}V_{00,00}^{l_{1}-1}
	\right]|\psi_{0}\rangle \nonumber
	\end{eqnarray}
\end{enumerate}


\end{document}